\documentclass[aps,pre,twocolumn,showpacs,superscriptaddress,groupedaddress]{revtex4}
\pdfoutput=1

\usepackage{hyperref}
\usepackage{color}
\usepackage{graphics,graphicx,epsfig,wrapfig}
\usepackage{epsf,epstopdf}
\usepackage{amssymb,amsfonts,amsmath}
\usepackage{enumitem}
\usepackage{ifthen}

\newcommand\mydots{\hbox to 0.8em{.\hss.\hss.}}
\newcommand{\beqn}{\begin{eqnarray}}
\newcommand{\eeqn}{\end{eqnarray}}
\newcommand{\beq}{\begin{equation}}
\newcommand{\eeq}{\end{equation}}

\newcommand{\abs}[1]{|#1|}

\newcommand{\sigmahat}{\hat{\sigma}}

\newcommand{\MM}[1]{{\color{black}#1}}

\begin{document}
\title{Trade-offs in delayed information transmission in biochemical networks.}
\author{F.~Mancini}
\email[]{fmancini@sissa.it}
\affiliation{International School for Advanced Studies (SISSA), Trieste, Italy}

\author{M.~Marsili}
\email[]{marsili@ictp.it}
\affiliation{The Abdus Salam International Centre for Theoretical Physics (ICTP), Trieste, Italy}

\author{A.~M.~Walczak}
\email[]{awalczak@lpt.ens.fr}
\affiliation{CNRS and Laboratoire de Physique Th\'{e}orique de l'\'{E}cole Normale Sup\'{e}rieure, Paris, France. }

\date{\today}
\linespread{1}

\begin{abstract}
In order to transmit biochemical signals, biological regulatory systems dissipate energy with concomitant entropy production. Additionally, signaling often takes place in challenging environmental conditions. In a simple model regulatory circuit given by an input and a delayed output, we explore the trade-offs between information transmission and the system''s energetic efficiency. We determine the maximally informative network, given a fixed amount of entropy production and delayed response, exploring both the case with and without feedback. We find that feedback allows the circuit to overcome energy constraints and transmit close to the maximum available information even in the dissipationless limit. Negative feedback loops,  characteristic of shock responses, are optimal at high dissipation. Close to equilibrium positive feedback loops, known for their stability, become more informative. Asking how the signaling network should be constructed to best function in the worst possible environment, rather than 
an optimally tuned one or in steady state, we discover that at large dissipation the same universal motif is optimal in all of these conditions.
\end{abstract}

\maketitle

\section{Introduction}

Cells respond to the current state of their environment by processing external signals through molecular networks and cascades. An external chemical stimulus is measured by receptors, which activate a series of biochemical reactions and lead the cell to produce an appropriate response. This response can be activating a gene or pathway, producing proteins that process the signal as in the case of sugar metabolism, result in motion such as in the case of chemotaxis, or initiating a cellular response such as apoptosis.
As we learn more about the structure of biochemical networks, we need to understand the functional role of their elements and connections. Yet regulation comes at  a cost, which imposes constraints on the form of these networks. Here we consider the limitations coming from thermodynamic constraints, caused by the cell's 
energy consumption, on the architecture of regulatory elements that best convey information about input signals to their outputs. We compare these most informative network structures to circuits that transmit the largest amount of information in unfavorable environmental conditions.

Despite the large complexity of biological regulatory networks, not all possible molecular regulatory circuits can be found in living organisms \cite{Alon2006}. One can ask whether the network architectures and parameter regimes are only shaped by the evolutionary history of these organisms, or whether there are also physical limits that constrain them. In the last years, a number of groups have explored different physical principles that could influence the parameter regimes and modes of regulation in living organisms (e.g. \cite{Hopfield1974, Ninio1975, Tostevin2006, Tostevin2008, Francois2004, Francois2007, saunders2009, Tkacik2008a,Mehta2009,Walczak2010, Dubuis2013, Tostevin2009, Tostevin2010, deRonde2010, Savageau1977, Scott2010, Aquino2014, Vergassola2007, Celani2010, Siggia2013}). One approach has been to calculate the limits that the intrinsic randomness in gene regulation imposes on information transmission between the input signal 
and its output responses, in networks of varying complexity \cite{Mugler2009, Tkavcik2009, Walczak2010, Tkavcik2012, Walczak2009, Mugler2010, Rieckh2014, Sokolowski2015}.  These studies showed which network architectures are optimal for information transmission and found that distinguishing different output states in general increases the transmitted information. They also pointed to the important trade offs between the information that the output has about the input and molecular costs.

The validity of the assumption that biochemical regulatory networks are maximally transmitting information between the concentrations of their input and output proteins was tested by Tkacik et al.~\cite{Tkavcik2008} in the case of the Bicoid morphogen gradient. Bicoid proteins regulate the expression of the hunchback gene in early fruit fly development. Using detailed measurements of the concentration and noise profiles of the Hunchback protein as a function of Bicoid concentration  \cite{Gregor2007, Gregor2007a}, the prediction for the probability distribution of the output concentration obtained by maximizing the flow of information was shown to match the experimental Hunchback distribution extremely well. In another combined experimental and theoretical study, Cheong et al.~\cite{Cheong2011} measured the amount of 
information transmitted to NF-kappa B controlled genes in the case of TNF stimulation. They showed how bottlenecks in this system reduce the amount of transmitted information compared to regulation via  multiple independent pathways. They argued that negative feedback, or information sharing between cells, can help transmit more information. The NF-kappa~B and ERK pathways were recently used to demonstrate that dynamical measurements of the response can transmit more information than static or one time readouts \cite{Selimkhanov2014}. 
Lastly, an information-theoretic approach was used in an experimental and numerical study to show the interdependence of stochastic processes controlling enzymatic calcium signaling  under different cellular conditions \cite{Pahle2008}.

Many of the current approaches to information transmission have looked at instantaneous information transmission \cite{Tkavcik2009, Walczak2010, Tkavcik2012}, or the rate of information transmission  \cite{Tostevin2009, Tostevin2010, Ronde2010, Ronde2012}. However, it has been argued that information transmission may be enhanced by dynamic biochemical readouts at multiple time points \cite{Selimkhanov2014} or when the regulatory response is at a delay relative to input signaling \cite{Nemenman2012}. 
Additionally, many biochemical networks function out of steady state, responding to inputs that are changing in time. Examples include the chemotactic response of bacteria or amoebas to nutrients or conversely to antibiotics.

Inspired by these observations, we previously studied the optimal circuits for transmitting information between an input and output read out with a fixed delay, in and out of steady state \cite{Mancini2013}. Delayed readouts are natural to most biochemical circuits, since sensing a signal requires production of the response, which takes time. For example, sensing an increased sugar concentration means the cell has to produce the enzyme to degrade it. We asked whether different readout delays correspond to different optimal circuits. We found that topologies of maximally informative networks correspond to commonly occurring negative feedback circuits irrespective of the temporal delay specified. Most interestingly, circuits functioning out of steady state may exploit non-equilibrium absorbing states to transmit information optimally and feedback can additionally increase information transmission. We found that there are many degenerate topologies that transmit similar information equally 
optimally - a degeneracy that will most likely be lifted by considering more detailed molecular models. 

The optimal solutions we found previously function strongly out of equilibrium, so they must consume energy. 
Since it has been experimentally shown \cite{Niven2008} that sensory systems may have evolved to reduce their energy expenditure,
we were interested in seeing how energetic constraints impact the form of the optimally informative solutions. This knowledge will prove useful when constructing artificial biochemical circuits \cite{Kalyanasundaram2010}, or engineering living organisms for energy production \cite{Lee2008}. 
The energy dissipated (or consumed) by a given network can be estimated by looking at the thermodynamics of its composite biochemical reactions. A completely reversible reaction does not consume energy. The reaction is in perfect equilibrium and the total free energy of the system is completely balanced. Irreversible reactions, such as certain steps of biochemical cascades, come at a cost to the cell, which has to prevent the back reaction from occurring. This cost can be estimated considering the flux balance of the network. The heat dissipated by the circuit is proportional to its rate of entropy production \cite{Crooks1998}. Tu et al. \cite{Tu2008} looked at entropy production in biochemical regulatory networks and experimentally showed that the flagellar motor switch of \textit{Escherichia coli} operates out of equilibrium, dissipating energy. A nonequlibrium allosteric model consistent with experimental results was proposed to explain how the switch operates with high sensitivity at a small energetic 
cost.

 Energetic cost has also been discussed in relation to cellular precision and the predictive power of the cell. The chemosensory system of \textit{E. coli} has been shown to dissipate energy in order to improve its adaptive speed and accuracy  \cite{Lan2012}. Reliable readout of input concentrations has also been bound by the entropy production rate  \cite{Mehta2012, Barato2013, Barato2014, Bo2015, Govern2014}. Others have reversed the perspective and shown that the minimum energy required for a biological sensor to detect a change in an environmental signal is proportional to the amount of information processed during the process \cite{Sartori2014}. In the case of the \textit{E. coli} chemosensory system, it was argued that 5\% of the energy consumed in sensing is determined by information-thermodynamic bounds, and is thus unavoidable \cite{Sartori2014}.  Becker et al. \cite{Becker2013} showed that short-term prediction in a sensory module is possible in equilibrium, but only up to a finite time interval. 
For longer times accurate prediction requires large dissipation. Lastly, the inability of systems to use all knowledge of past environmental fluctuations to predict the future state has been directly linked to dissipation \cite{Still2012}.

We want to see how the structure of optimal networks for information transmission changes if we impose a penalty on the entropy production of the system. In order to investigate the non-equilibrium nature of biochemical circuits that are optimal for delayed information transmission, we choose to study a  simple binary model of a regulatory circuit that allows us to focus on the regulatory logic at small computational costs. 
Within this model we consider two interacting elements of biochemical regulatory networks (e.g. proteins and genes, elements of two component signaling systems, sugars and enzymes) that take on binary states (on or off) and evolve in continuous time. This simplification allows us to develop an efficient formalism for calculating information transmission at different readout delays and consider the connection between dissipation and different readout times. In the limit of infinite dissipation rates, we recover the 
previously obtained results \cite{Mancini2013}. For finite, non zero dissipation rates, back reactions decrease the information transmission until it goes to zero for systems close to equilibrium. However, when feedback is allowed, networks are able to transmit almost 1 bit of information at no cost.

Optimizing biochemical networks for information transmission assumes that the circuit and its environment have coevolved to best match their statistical properties. For many networks this is a valid assumption. However,  other networks function in a wide variety of variable conditions. To study what kind of network is best adapted to function in adverse environments we combine a game-theoretic maximin approach with the framework of information theory. We ask what system will maximally transmit information even when presented by the environment with the worst possible initial state - the one that aims at minimizing information at all time delays. Interestingly, we find that, even if the amount of transmitted information is inevitably smaller, the structure of the optimal circuits is the same as when the environment has no detrimental effect and the system is able to optimize its initial condition.

Game-theoretic approaches have been used to robustly design biochemical networks and to devise biomimicking algorithms. Given environmental disturbances and uncertainty about the initial state, minimax strategies were used to match therapeutic treatment to a prescribed immune response 
\cite{Chen2008}, and to make a stochastic synthetic gene network achieve a desired steady state \cite{Chen2009}. 
The adaptive response in bacterial chemotaxis has been interpreted as a maximin strategy that ensures the highest minimum chemoattractant uptake for any profile of concentration \cite{Celani2010}.

In the first section of this paper we discuss the effect of energetic constraints on information transmission at a time delay. We consider the case in which the system is at steady state and the signal up-regulates (or down-regulates) the response with and without feedback. In the second section we investigate how the system counteracts the worst possible initial condition presented by the environment in order to transmit as much information as possible.
We finish with a discussion of our results and their interpretation in terms of biochemical regulatory networks.  

\section{Information transmission with energy dissipation}
\label{sec:diss-model}

\begin{figure}[!ht]
\includegraphics[scale=0.6]{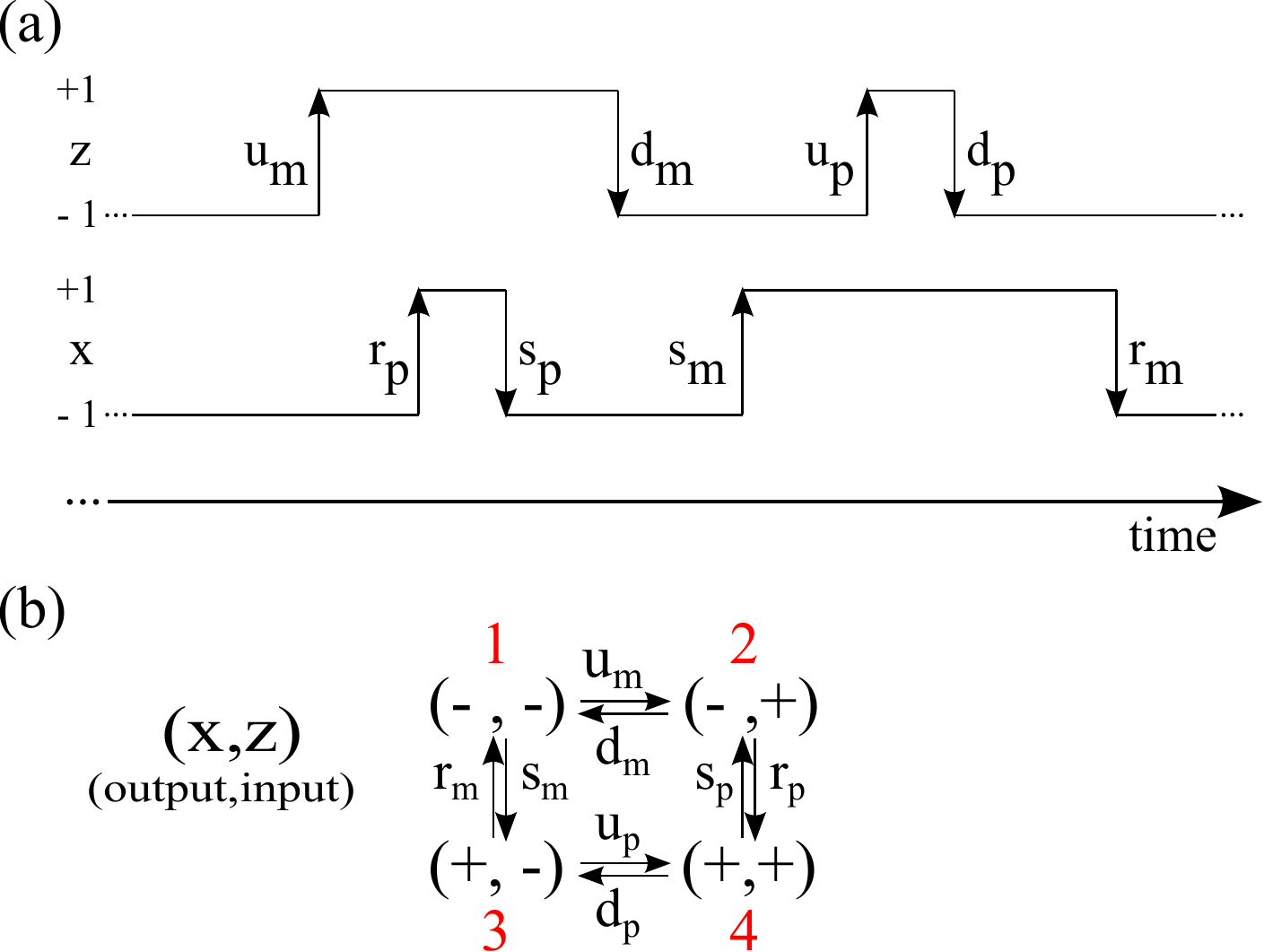} 
\caption{(a) \label{fig:diagram}  Time evolution of the random variable $z_t$, which models a biochemical input transitioning from/to a down-regulated state ($-1$) to/from an up-regulated state ($+1$), with rates $\{u_m,u_p\}$/$\{d_m,d_p\}$, respectively. The random variable $x_t$ models activation ($+1$) or deactivation($-1$) of a biochemical output: it is regulated by $z$, with which it aligns (`activation', or up-regulation) with rates $r_m$ or $r_p$ or anti-aligns (`repression', or down-regulation) with rates $s_m$ or $s_p$. The subscripts $m$ and $p$ in the rates account for the state of the other variable, that is $-1$ and $+1$, respectively. 
(b) \label{fig:drawing} The four network states, with corresponding transition rates given in (a).}
\end{figure}

To focus on the tradeoffs between the ability of the network to transmit information and the energetic cost of the biochemical reactions that make up this network, we study the simplest model of regulation that allows us to focus on the logic of the reactions.  Our simplified network (see Fig.~\ref{fig:diagram}) consists of two binary elements that describe either a transcription factor protein regulating a gene, or a signaling molecule activating/downregulating an enzyme or receptor. The first element of the network describes the input $z$ and can be associated with the state of a receptor, signaling molecule or transcription factor that responds to the external conditions. For example, it can describe the presence or absence of a sugar source in metabolism or phosphorylation of the histidine kinase in a two component  signaling system. The output $x$ describes the final outcome of the network, such as the gene that produces the response protein to the external signal. In the examples given above, it 
corresponds to the enzyme that digests the sugar or expression of the  target gene by the response regulator. 
Both of these elements can be found in the active ($x,z=+1$) or inactive ($x,z=-1$) states.
If the described element is a continuous variable (e.g. protein concentration), the binary approximation is equivalent to taking very steep regulatory functions, such that the concentration is well described by two states: below and above the threshold. 

This two component system can be found in one of four states: $(x, z) \in \{ (-, -), (-, +), (+, -), (+, +)\}$, corresponding to  both elements inactive, the input active --- output inactive and vice versa, and both elements active. 
The input $z$ up/down regulates $x$ with rates $r_m (r_p)$ and $s_m (s_p)$, defined in Fig.~\ref{fig:diagram},  that depend on the state of the input ($m=-,p=+ $). 
The state of the system is defined by the {\MM{conditional probability $P(x_t,z_t,t|x_0,z_0,0)$ to find the system in state $(x_t,z_t)$ at time $t$, conditional on the state $(x_0,z_0)$ at time $t=0$. This conditional probability distribution can be arranged in a $4\times 4$ matrix, and its evolution is described by the master equation}}
\beq
\label{eqn:master_eq}
\partial_t P= -{\cal L} P
\eeq
{\MM{where the $4\times 4$ transition matrix ${\cal L}$ is defined in terms of the rates depicted in the diagram in Fig.~\ref{fig:diagram}  (see Appendix \ref{app0}). The central quantity we shall be interested in is the joint probability distribution of the state $x_t$ of the output at time $t$ and the state $z_0$ of the input at time $0$. We shall use the shorthand 
\beq
\label{eq:Pxtz0_sum}
P(x_t,z_0) = \sum_{z_t,x_0=\pm 1} P(x_t,z_t,t|x_0,z_0,0)P_0(x_0,z_0)
\eeq
where $P_0(x_0,z_0)$ is the probability distribution of the system at the initial time.}}

We are interested in finding the network topologies that are optimal for information transmission over a fixed time scale $\tau$. Specifically we want to maximize the mutual information between an input signal at an initial time $z_0$, and the output of the network which is read out at a later time, $x_t$:
\beq
\label{eq:rescMI}
\mathcal{I}(\tau)=I[x_{t=\tau/\lambda},z_0] 
\eeq
 over the rates of the biochemical reactions $\mathcal{L}$ of the regulatory network, where:
 \beq
  \label{eq:MI}
 I[x_t, z_0]=\sum_{x_t,z_0} P(x_t,z_0)\log\frac{P(x_t,z_0)}{P(x_t)P_0(z_0)},
 \eeq
{\MM{ and $P_0(z_0) =\sum_{x_0} P_0(x_0,z_0)$, $P(x_t)=\sum_{z_0}P(x_t,z_0)$. }}
  We will measure the time $t=\tau/\lambda$ between the signal and the delayed read-out in units of the natural timescale of the problem -- the relaxation time $\lambda^{-1}$, calculated as the inverse of the minimal non-zero eigenvalue of $\mathcal{L}$. Previously we found the networks that are best suited for transmitting information at a delay and discovered that they correspond to systems that function out of equilibrium. 
 
 For this reason we are interested in posing the same question, but taking into account energy constraints. 
 We thus constrain the energy $\dot{Q}$ dissipated per unit time into an external medium at temperature $T$ that is in contact with our system. $\dot{Q}$ is related to the thermodynamic entropy production rate $\sigma$,  $\dot{Q} = k_BT \sigma$, where $k_B$ is the Boltzmann constant \cite{Crooks1998, Seifert2012}. In steady state the thermodynamic entropy production rate takes the form
\beq
\label{eq:sigma}
\sigma= \sum_{i,j}P^{\infty}_iw_{ij}\log\frac{w_{ij}}{w_{ji}},
\eeq
where, in terms of the shorthand $i,j=(x_t,z_t)$ to denote the states, $w_{ij}$ is the transition rate from state $i$ to $j$ and $P^{\infty}_i$ is the steady state probability distribution for state $i$. 
 
In order to intuitively understand the expression in Eq.~\ref{eq:sigma} we link it to the non equilibrium properties of the system. In steady state the master equation satisfies
\beq
\label{eq:StatState}
P^{\infty}_iw_{ij}- P^{\infty}_jw_{ji} = \pm J
\eeq
with the $+$ ($-$) sign that holds for all pairs of states where $i$ follows $j$ in the clockwise direction in Fig.~\ref{fig:drawing}, and $J$ is the steady state current. The detailed balance condition 
\beq
\label{eq:DB}
P^{\infty}_iw_{ij}- P^{\infty}_jw_{ji} = 0 \quad \forall i,j,
\eeq
is a special case of Eq.~\ref{eq:StatState} where $J=0$. 
In terms of the current defined in Eq.~\ref{eq:StatState},  the steady state entropy production rate in Eq.~\ref{eq:sigma} becomes
 \beq
\sigma= J \log\frac{w_{12}\, w_{24} \, w_{43} \, w_{31}}{w_{21}\, w_{13} \, w_{34} \, w_{42}}
\eeq 
 (see Appendix \ref{app1} for derivation).
In order to maintain a non-equilibrium steady state ($J\neq 0$) the system has to dissipate energy at rate $k_BT\sigma$.

We are interested in solving the problem of finding the best network design that can perform a maximally informative delayed readout given a limited and fixed amount of $k_BT\sigma$ units of energy per unit time. This question can be addressed quantitatively by introducing a Lagrange multiplier $l$ that constrains the energy cost of the transmitted information and maximizing the functional
\beq
\label{ }
\mathcal{I}(\tau)-l\frac{\sigma}{\lambda \log 2} = \mathcal{I}(\tau) - l\sigmahat
\eeq
over the circuit's reaction rates, $\mathcal{L}$. We rescale the rate of energy dissipation, $\sigma$, by the constant $\lambda \log 2$ and call it $\sigmahat$, in order to express both information and entropy production in bits and to measure time in units of the characteristic timescale $1/\lambda$.

For $l=0$ the constraint on the dissipated energy does not enter the optimization and one recovers the results found without imposing energetic constraints ($\sigmahat=\sigma=\infty)$  \cite{Mancini2013}. In this limit the system is driven out of equilibrium and at least one of the rates vanishes. At the other extreme, when $l=\infty$, any deviation from equilibrium is severely punished  and we expect to find the system in equilibrium.

Some intuition about the optimal solutions can be gained before embarking on detailed calculations. In general we can write the probability distribution as $P(x_t,z_0)=(a+bx_t+cz_0+ \mu x_tz_0)/{4}$. The symmetry between the on and off states, 
$P(+,+)=P(-,-)$ and $P(+,-)=P(-,+)$,  implies $b=c=0$ and normalization $\sum_{\rm i,j} P(i,j)=1$ gives $a=1$. Therefore $P(x_t,z_0)$ has to be of the form 
\beq
\label{eq:Pxtz0_mu}
P(x_t,z_0)=\frac{1+\mu x_tz_0}{4}.
\eeq
Eq.~\ref{eq:Pxtz0_mu} means $P(x_t)=P(z_0)=1/2$, which independently maximizes the entropy  of the input, $H[z_0]$, and output distribution, $H[x_t]$, where $H[y]=-P(y) \log P(y)$. With this form for $P(x_t,z_0)$ the mutual information in Eq.~\ref{eq:MI} becomes:
\beq
\label{eq:MI_vs_mu}
I=\frac{1+\mu}{2}\log(1+\mu)+\frac{1-\mu}{2}\log(1-\mu),
\eeq
where in general $\abs{\mu}\leq 1$. The symmetry of the system results in a degeneracy of solutions, which we break by setting one of the input flipping rates to a fixed value $r=1$. 
With this choice, the allowed range of $\mu$ is  $[0, 1]$, and information is a monotonically increasing function of the ``effective magnetization'' $\mu$ and is maximized for $\mu=1$ giving $I=1$ bit. We compute $\mu$ explicitly for specific models in the following sections.

\subsection{Simplest model}
\label{subsec:diss_no_feedback}

\begin{figure}[!ht]
\hspace{- 0.1\linewidth}
\includegraphics[scale=0.7]{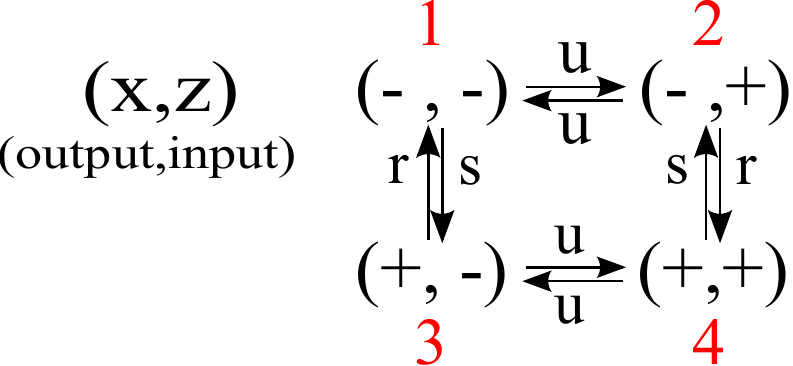}
\caption{\label{fig:drawing_modelA} The four network states, with corresponding transition rates, in the simplest case where input $z$ can either up or down-regulate the output $x$ but there is no feedback. The input $z$ switches with the same rate regardless of the state of the output $x$.}
\end{figure}

First we consider the simplest case depicted in Fig.~\ref{fig:drawing_modelA}, where we set all the rates for flipping of the input $z$ to be equal $u_p = u_m = d_p = d_m \equiv u$, but allow the rates for the output $x$ to be different if the output is aligning with the input $r_p = r_m \equiv r$ or it is anti-aligning $s_p = s_m \equiv s$. This models allows $z$ to activate (or repress) $x$ with rate $r$($s$), respectively, but does not allow for feedback since the flipping rate of the input does not depend on the state of the output. We diagonalize the rate matrix $\cal{L}$ for this model analytically and find the eigenvalues to be $\{0, 2u, 1+s, 1+s+2u\}$ (see Appendix \ref{appA_diagL} for details).

We can express the mutual information explicitly in the form given by Eq.\ref{eq:MI_vs_mu}, with 
\beq
\mu = (1-s) \frac{(1+s+2u)e^{-\frac{2u}{\lambda} \tau} -4u e^{-\frac{(1+s)}{\lambda}\tau}}{(1+s)^2 - 4u^2},
\label{eq:mumodelA}
\eeq
where time is rescaled by the smallest nonzero eigenvalue $\lambda$, as specified in Eq.~\ref{eq:rescMI}.

The rescaled entropy production rate (Eq.~\ref{eq:sigma}) becomes
\beq
\label{eq:sigma_no-feedback}
\sigmahat = \frac{(1-s) u \log \frac{1}{s}}{\lambda (1+s+2u)}.
\eeq
Given that the smallest nonzero eigenvalue can be either $\lambda = 2u$ or $\lambda = 1+s$, we can define the quantity $\gamma = ({1+s})/({2u})$ and distinguish two regimes: $\gamma \geq 1$ in which the output changes on faster timescales than the input  and $\gamma  \leq 1$, where the input changes more quickly than the output. In general, for each set of rates, the two eigenvalues must be compared and the value of $\lambda$ (and thus $\gamma$) determined.

\subsubsection{Numerical results:} 

To get an idea about the behavior of the system we will first solve the optimization problem numerically and then interpret the results in terms of the limiting cases. 
For each readout delay $\tau$ and entropy production rate $\sigmahat$, we look for rates that maximize $\mathcal{I}(\tau)$ (given by Eqs.~\ref{eq:MI_vs_mu} and \ref{eq:mumodelA}) while constraining the rates at fixed $\sigmahat$ (given by Eq.~\ref{eq:sigma_no-feedback}). 

\begin{figure*}[!ht]
\hspace{\linewidth}
\includegraphics[scale=0.7]{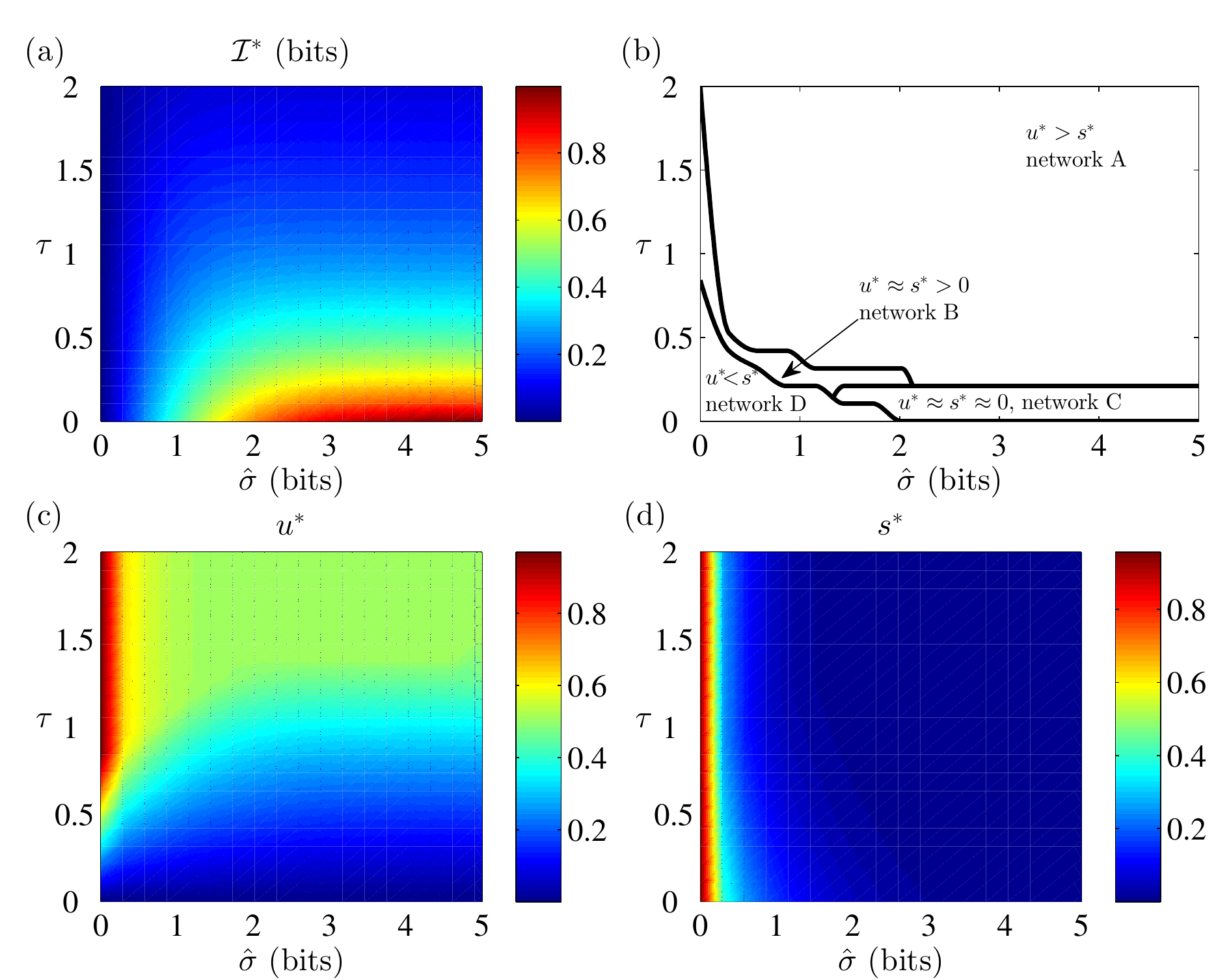}\\
\vspace{0.05\linewidth}
\includegraphics[scale=0.6]{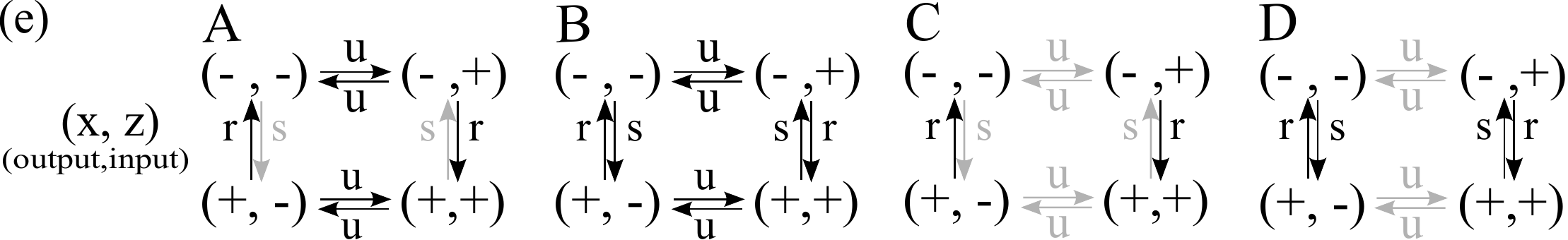}
\caption{
(a)\label{fig:contourMaxInfo} Contour plot of optimal mutual information $\mathcal{I}^*$ as function of the readout delay $\tau$ and entropy production rate $\sigmahat$.
(b) \label{fig:phaseDiagram_modelA} Phase diagram in the $(\sigmahat,\tau)$ plane of the optimal network topologies A, B, C, D (sketched in panel (e)).
(c) \label{fig:contourOptu} Contour plot of optimal rate $u^*$ as function of the readout delay $\tau$ and entropy production rate $\sigmahat$.
(d) \label{fig:contourOpts} Contour plot of optimal parameter $s^*$ as function of the readout  delay $\tau$ and entropy production rate $\sigmahat$.
(e) \label{fig:optimalNetworks_modelA} Sketch of optimal network topologies A, B, C, D.}
\end{figure*}

The maximal mutual information values (capacities) of the optimal networks display an intuitive behavior as functions of the dissipated energy and time delay of the readout.  The mutual information between the input and output of the optimal network decreases with the time delay of the input readout for all values of dissipation (see Fig.~\ref{fig:contourOptu}~a), as the network decorrelates. Allowing the system to dissipate more energy 
increases its capacity to transmit information. Above a certain value of dissipated energy the capacity plateaus and reaches the same value we observed if we did not constrain dissipation \cite{Mancini2013}. The value of the this plateau decreases with an increase of the time delay $\tau$ of the readout (see section~\ref{subsubsec:sigma_very_large_modelA} for a functional dependence). The transmitted information decreases to zero linearly with dissipation for all readout delays, $\mathcal{I}^* \sim \frac{c(\tau)^2 \sigmahat}{2 {\log 2}}$, where $c(\tau)$ 
is a $\tau$ dependent constant derived in section \ref{subsubsec:sigma_very_small_modelA}.  
Naturally, the capacities for systems that can dissipate a lot of energy are much larger than those with large energy constraints. However at small time delays the rate of decay of the capacity with time delay is larger for circuits that function far out of equilibrium than those that are close to equilibrium (see section \ref{subsubsec:tau_very_small}). 

In Figures~\ref{fig:contourOptu}~C and \ref{fig:contourOptu}~D we plot the values of the rate constants of the optimal networks that result in the capacities plotted in Fig.~\ref{fig:contourOptu}~a. We see that similarly to the capacity values the optimal rates are continuous. To gain a better idea about the network topologies that give optimal networks we have used the rates to broadly classify the circuit topologies in the phase diagram in Fig.~\ref{fig:phaseDiagram_modelA}~b with the topologies defined in Fig.~\ref{fig:phaseDiagram_modelA}~e. In the limit of large dissipation we recover the results we obtained previously \cite{Mancini2013}: in the optimal circuit at large readout delays the flipping of the output is governed by an irreversible fast reaction with rate $r$ fixed to $1$  (the back reaction is forbidden $s^*=0$). The output follows the state of the input and the change in the input is described by a 
reversible slower reaction with rate $u^*<r^*$ (network A in Fig.~\ref{fig:contourOptu}~e). For shorter delays the flipping rate of the input decreases, causing the capacity to increase. As $\tau \rightarrow 0$, $u \rightarrow0$ and we obtain two separate subnetworks with a fixed input in which the output changes quickly to follow the input (model C in Fig.~\ref{fig:contourOptu}~e). 

At large readout delays, the equilibrium solution at $\sigmahat \rightarrow 0$ is very similar to the non-equilibrium one, but now detailed balance must always be obeyed. The detailed balanced condition imposes that the output change is completely reversible and now $s^*\neq0$. At $\sigmahat=0$ the forward and back reactions are completely balanced with $s^*=r$ (network B in Fig.~\ref{fig:phaseDiagram_modelA}~e). Additionally, the input changes on the same very fast timescale $u^*\approx r=1$, faster than for large $\sigmahat$. Not surprisingly this essentially randomly flipping equilibrium circuit at large delays is not able to reliably transmit information, 
and ${\cal I} \approx 0$. For short time delays, similarly as in the large $\sigmahat$ limit, $u^*\rightarrow 0$ and we obtain two subcircuits with the output flipping back and forth at the same rate $s^*=r$, at $\sigmahat=0$ (network D in Fig.~\ref{fig:phaseDiagram_modelA}~e). Allowing for small amounts of dissipation breaks detailed balance and decreases the rate of the output's back reaction ($s^*<1$), so that the output is more likely to be in the same state as the input. 

In summary, network C that has a fixed input, which is followed by the output on fast timescales, is the most informative solution. The capacity of this system is reached at finite values of $\sigmahat$, and does not  increase further as $\sigmahat \rightarrow \infty$. This topology is optimal for a wide range of $\sigmahat$, with the back reaction rate $s$ continuously increasing as the constraints on dissipation impose solutions closer to equilibrium, until network D with the randomly flipping outputs is reached. At small time delays the optimal solution always keeps the input fixed and adjusts the state of the output to the input ($2u\leq r$). But for large $\tau$ the input will change ($2u \sim r$) and the amount of energy that can be dissipated controls whether the output simply follows the input (network A in Fig.~\ref{fig:phaseDiagram_modelA}~e), or is forced to switch independently (network B in Fig.~\ref{fig:phaseDiagram_modelA}~e). Information can therefore be lost both in circuits where the output 
does not have the energy to follow the input (network D) and in circuits where the 
input decorrelates with time (network A), or both of these scenarios apply (network B).

\begin{figure}[!ht]
\hspace{- 0.1\linewidth}
\includegraphics[scale=0.5]{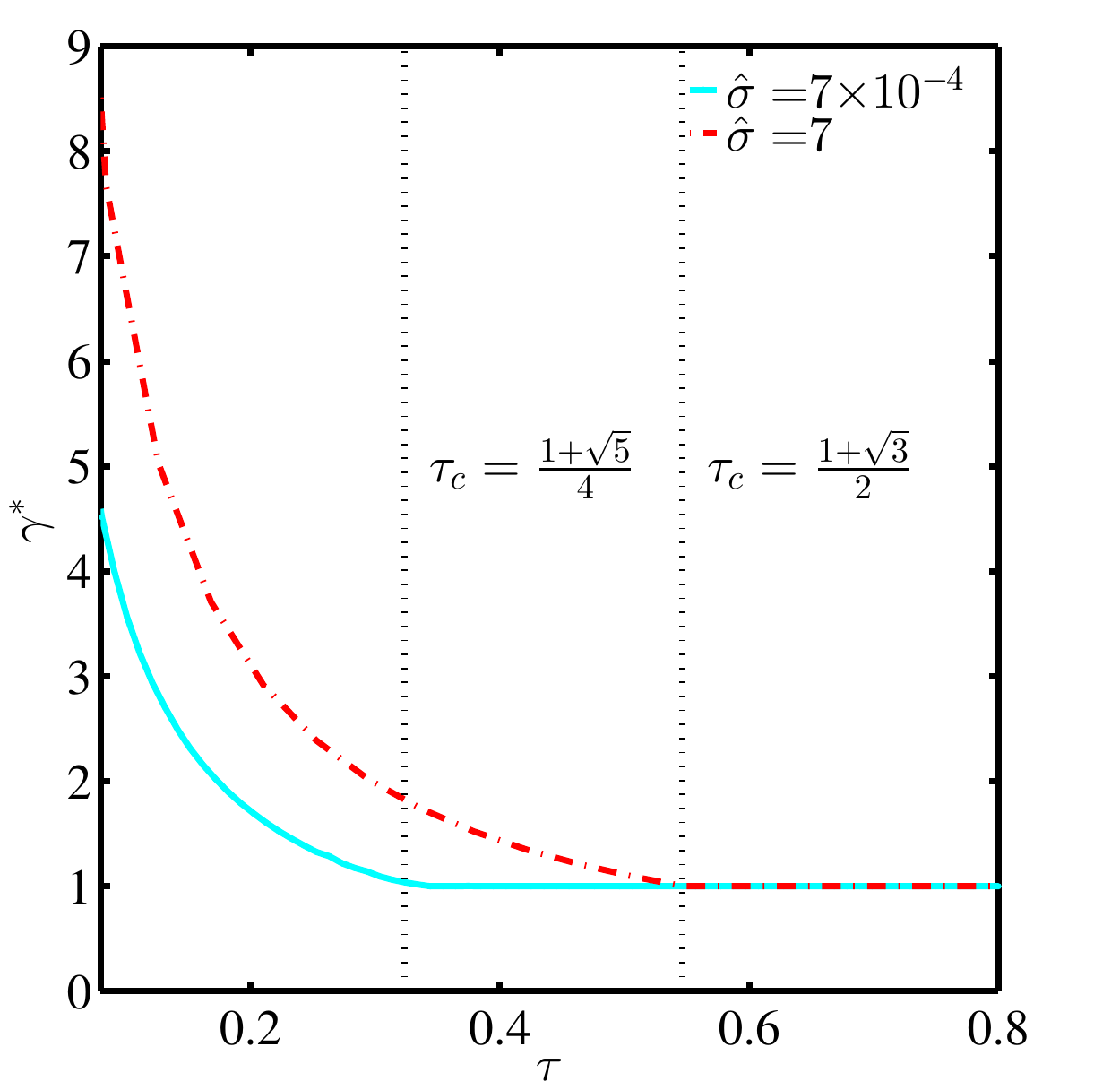}
\caption{\label{fig:checkTauC} The quantity $\gamma=(1+s)/(2u)$, is the ratio of two smallest nonzero eigenvalues, corresponding to the output and the input timescale, respectively. The optimal value $\gamma^*$ is shown as a function of delay $\tau$, in the two limits of very small and very large entropy production $\sigmahat$ (measured in bits). The time $\tau_c$ after which the output timescale matches the input timescale ($\gamma^*=1$) decreases with dissipation from ${1+\sqrt{3}}/{2}$ when $\sigmahat\rightarrow \infty$ to ${1+\sqrt{5}}/{4}$ when $\sigmahat \rightarrow 0$.}
\end{figure}

Lastly, one can interpret the optimal circuits in terms of the relaxation rate of the system (smallest nonzero eigenvalue). The ratio of the two potentially smallest eigenvalues $\gamma$ is given by $({1+s})/({2u})$ -- the ratio of the output and the input switching rates. Fig.~\ref{fig:checkTauC} shows the optimal value of $\gamma^*$, as a function of the delay $\tau$, in the limit of small entropy production ($\sigmahat=0.0007$ bits) and of large entropy production ($\sigmahat=7$ bits). As noted before, for small time delays optimal circuits are those where the input changes more slowly than the output ($\gamma^*>1$), for all values of dissipation. However for large $\tau$, we define a certain value $\tau_c$, at which the input and the output timescales match in optimal circuits, with $\gamma^*=1$. The value of $\tau_c$ corresponds to the optimal rate of input flipping $u^*$ reaching a constant value and depends on the rate of dissipation. 
For  $\sigmahat\ll1$, $\tau_c=({1+\sqrt{5}})/{4}$ and $u^*\sim1$ (see section~\ref{App:subsubsec:sigma_very_small_modelA} for a derivation). For large 
dissipation rates $\sigmahat\gg1$,  
this delay increases, $\tau_c=({1+\sqrt{3}})/{2}$, and the rate of change of the input decreases to $u^*=0.5$ (see section~\ref{subsubsec:sigma_very_large_modelA}). At large delays $\tau$, matching the input and output switching rates allows the system to transmit more information. This matching of timescales is possible at $\tau>\tau_c>0$ even if the system cannot dissipate energy (small $\sigmahat$). Finally, the optimal solution always is in the $\gamma^* \geq 1$ limit, where the input changes more slowly than the output. 

Having understood the general behavior of the capacity of this model, we can exploit its simplicity to obtain precise analytical scaling results in the limits of small and large delays and dissipation.

\begin{figure*}[!ht]
\includegraphics[scale=0.7]{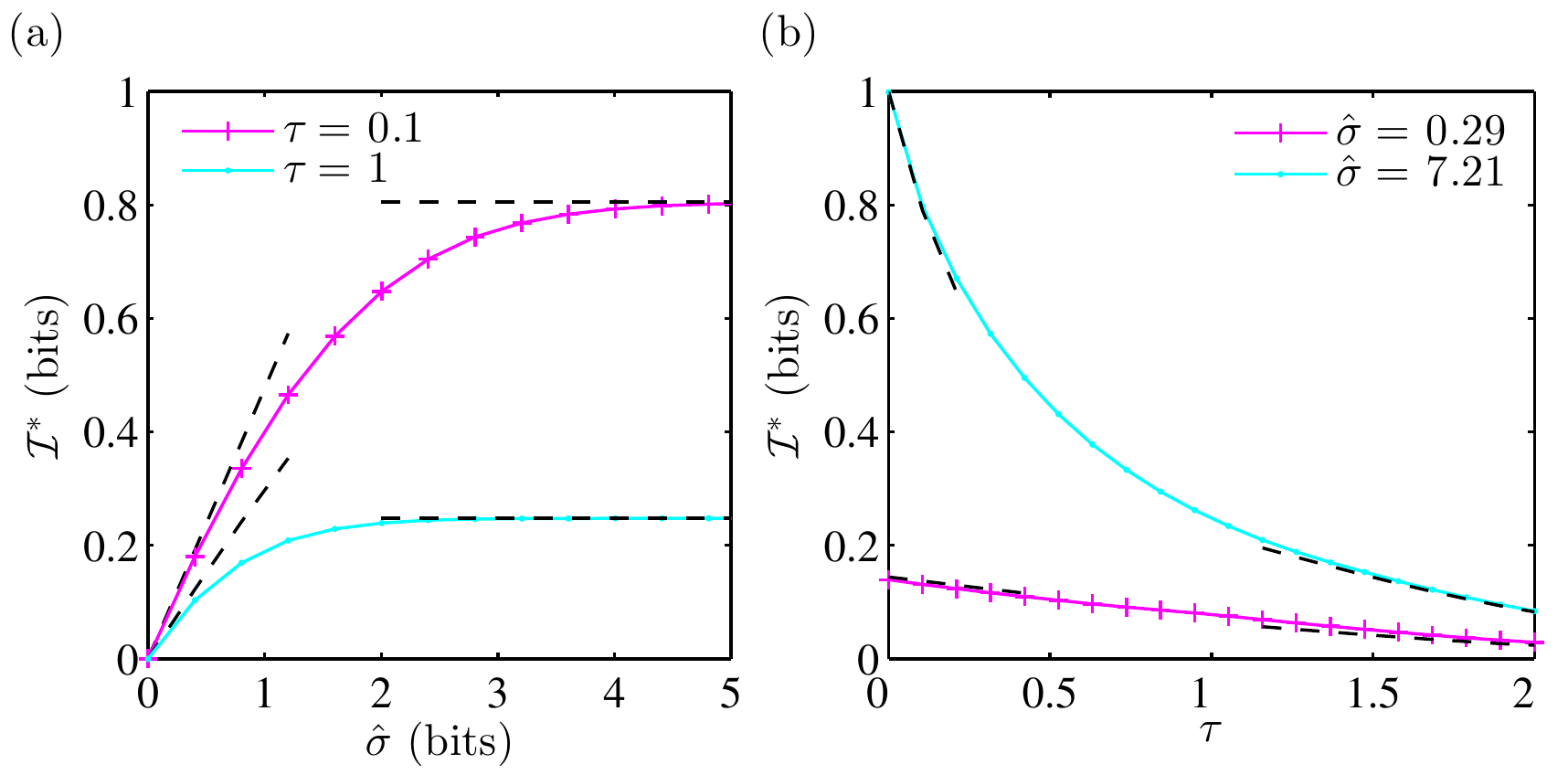}
\caption{\label{fig:finalAnalyticalNumericalChecks} Comparison of the analytical (dashed lines) and numerical solutions (solid lines) for optimal mutual information $\mathcal{I}^*$. In panel (a) the behavior for small values of entropy production $\sigmahat$ is shown for $\tau=0.1$ and for $\tau=1$ as presented in section~\ref{subsubsec:sigma_very_small_modelA}. In panel (b) the dependence on $\tau$ is represented for $\sigmahat=0.29$ bits and $\sigmahat=7.21$ bits as presented in sections~\ref{subsubsec:sigma_very_small_modelA} and \ref{subsubsec:sigma_very_large_modelA}.}
\end{figure*}

\subsubsection{Limit \texorpdfstring{$\tau=0$}{tau=0}}
The simplest case is that of instantaneous readout, $\tau=0$, where the effective magnetization $\mu$ is
\beq
\mu = \left(\frac{1-s}{1+s}\right)\frac{\gamma}{\gamma+1}.
\eeq
We can formally rewrite Eq.~\ref{eq:sigma_no-feedback} as:
\beq
\label{eq:expr_s}
s = \exp\left[-2\beta(\sigmahat, \gamma) \sigmahat \right],
\eeq
where $\beta(\sigmahat, \gamma)$ is in general a nonlinear function $\sigmahat$ and $\gamma$. This form agrees with the numerical results for $s^*$ that shows a strong decay with $\sigmahat$ (Fig.~\ref{fig:contourOpts}).

As we know from our numerical exploration, in the $\tau=0$ limit the capacity strongly depends on the value of $\sigmahat$. First we can explore the limit of large dissipation, where we know from our previous work (and from the results presented in Fig.~\ref{fig:contourOpts}) that $s^*$ is small. In this limit Eq.~\ref{eq:expr_s} simplifies (Eq.~\ref{eq:sigma_no-feedback} is explicitly solved for $s$) and $\beta$ is a function of only $u$ and $\lambda$, not of $\sigmahat$. To find $\gamma^*$ that maximizes $\mu$, we  exploit the parametrization of $s$ in Eq.~\ref{eq:expr_s} to write
\beq
\label{tau0mu}
\mu = \tanh \left(\beta(u, \lambda) \sigmahat \right) \frac{\gamma}{1+\gamma}.
\eeq
At fixed but large $\sigmahat$ the largest value of $\mu$ is always achieved for $\gamma^*=\infty$. This means the output changes on faster timescales than the input and the smallest eigenvalue is $\lambda=2u$. More precisely, as in the dissipation-less case \cite{Mancini2013}, the optimal rate is $u^*=0$ at $\tau=0$. 

To find $s^*$, we substitute the parametrization in Eq.~\ref{eq:expr_s} for $s$ into Eq.~\ref{eq:sigma_no-feedback} with $\lambda=2u$ and obtain at fixed $\sigmahat$:
\beq
\label{eq:gamma_vs_beta_firstcase}
1 + \frac{1}{\gamma} = \beta \tanh(\beta \sigmahat).
\eeq 
Since $\gamma^* = \infty$,  $\beta^*$ must satisfy $\beta^* \tanh(\beta^* \sigmahat)=1$. For large dissipation rates, $\beta^* \sim 1$, $s^*\sim e^{-2\sigmahat}$ is exponentially small as we had assumed and $\mu^* \sim \tanh(\sigmahat) \sim 1$. This results in the optimal information $\mathcal{I}^* \sim 1$ bit. 

For small dissipation rates, the general expression in Eq.~\ref{eq:expr_s} holds, where $\beta$ is a nonlinear  function of $\sigmahat$, $\beta(\sigmahat)$. Eqs.~\ref{tau0mu} and \ref{eq:gamma_vs_beta_firstcase} and the arguments presented above still hold, resulting in a maximum $\mu(\tau=0)$ when $\gamma^*=\infty$ and $u^*=0$. For small dissipation rates and taking $\gamma^*=\infty$, Eq.~\ref{eq:gamma_vs_beta_firstcase} becomes $\beta^*(\sigmahat) \sim 1/\sqrt{\sigmahat}$, and Eq.~\ref{tau0mu} results in the effective magnetization $\mu\sim \beta^*(\sigmahat) \sigmahat \sim \sqrt{\sigmahat}$.  Finally, the optimal mutual information goes to $0$ linearly with the rescaled dissipation $\mathcal{I}^* \approx (\mu^*)^2/2 \simeq \sigmahat/2$ bits.

\subsubsection{Limit \texorpdfstring{$\tau\ll1$}{tau very small}}
\label{subsubsec:tau_very_small}

The results from the $\tau=0$ limit serve as a basis for considering  the scaling of the mutual information for small, but finite $\tau\ll1$. Since  $\gamma^* $ diverges at  $\tau= 0$, we assume that for $\tau \rightarrow 0$  the smallest eigenvalue is still $2u$ and $\gamma^*>1$. We will also use the generalized nonlinear parametrization of $s$ in Eq.~\ref{eq:expr_s} with $\beta(\sigmahat)$, also in the small dissipation regime, as we did for $\tau=0$.

In the small  dissipation limit $\sigmahat\ll1$, Eq.~\ref{eq:gamma_vs_beta_firstcase} becomes:
\beq
\label{eq:beta_firstcase}
\beta \sim \frac{1}{\sqrt{\sigmahat}}\sqrt{\frac{\gamma+1}{\gamma}}.
\eeq
Using Eq.~\ref{eq:expr_s} the effective magnetization at fixed $\sigmahat$ is a function of only $\gamma$ and $\sigmahat$
\beq
\label{eq:mu_smalltau_smallsigma}
\mu =  \sqrt{\sigmahat}\sqrt{\frac{\gamma+1}{\gamma}}\frac{\gamma}{\gamma^2-1} [(\gamma+1)e^{-\tau} - 2e^{-\gamma\tau}].
\eeq
We maximize the effective magnetization with respect to $\gamma$, $d \mu/d\gamma=0$, and assume the scaling $\gamma^* \simeq \frac{a_0}{\tau} + b_0 + c_0 \tau$. Solving the resulting equations for the coefficients in orders of $\tau$ (see Appendix~\ref{appA_tausmall} for details), the optimal effective magnetization is 
\beq
\label{eq:mu_smalltau_smallsigma_final}
\mu^* \sim \sqrt{\sigmahat}(1 + A_0 \tau),
\eeq
where $A_0=-0.24\mydots$ is computed exactly in Appendix~\ref{appA_tausmall}.

In the large dissipation limit $\sigmahat \gg1$, Eq.~\ref{eq:gamma_vs_beta_firstcase} becomes
\beq
\beta \simeq \frac{\gamma+1}{\gamma}.
\eeq
Using Eq.~\ref{eq:expr_s}, $\lambda=2u$ and the fact that in this limit $s\rightarrow 0$, the effective magnetization in Eq.~\ref{eq:mumodelA} is
\beq
\label{eq:mu_smalltau_largesigma}
\mu =  \frac{\gamma}{\gamma^2-1} [(\gamma+1)e^{-\tau} - 2e^{-\gamma\tau}].
\eeq
Assuming $\gamma^*\simeq \frac{a_{\infty}}{\tau} + b_{\infty} + c_{\infty} \tau$, the analogous calculation to the small dissipation limit results in the maximized effective magnetization 
\beq
\label{eq:mu_smalltau_largesigma_final}
\mu^* \sim 1 + A_{\infty} \tau + B_{\infty} \tau^2,
\eeq
where $A_{\infty}=-0.63\mydots$ and $B_{\infty}=0.23\mydots$ are computed exactly in Appendix~\ref{appA_tausmall}.

Summarizing, in the small dissipation limit we find $\mathcal{I}^* \approx (\mu^*)^2/2 \sim \sigmahat(1+2A_0\tau)/2$ bits, a linear scaling of the information both with dissipation and with readout delay. In the large dissipation limit the information is independent of the dissipation and $\mathcal{I}^* \rightarrow 1$ bit, as $\mu^*$  tends to one quadratically in the delay, as given by Eq.~\ref{eq:mu_smalltau_largesigma_final}. 
This scaling behavior is compared with numerical results in Fig.~\ref{fig:finalAnalyticalNumericalChecks}.


\subsubsection{Limit \texorpdfstring{$\sigmahat\ll1$}{sigmahat very small}}
\label{subsubsec:sigma_very_small_modelA}

The scaling at small dissipation with $\sigmahat$ for all $\tau$ is obtained by noting from Eq.~\ref{eq:sigma_no-feedback} that in this limit $s \rightarrow 1$. In the regime where $\lambda=2u$, this behavior is clear. If $\lambda=1+s$, small $\sigmahat$ could also be obtained setting $u\simeq0$, yet this would mean that $2u<1+s$, and $\lambda=2u$. So the only consistent solution demands $s \rightarrow 1$. 

We set $s = 1-\epsilon$ and expand  Eq.~\ref{eq:mumodelA} to leading order in $\epsilon$
\beq
\label{eq:mu_eps_firstcase}
\mu \simeq \frac{\epsilon}{2}\gamma \frac{(1+\gamma)e^{-\tau} - 2e^{-\gamma\tau}}{\gamma^2-1},
\eeq
and similarly Eq.~\ref{eq:sigma_no-feedback} 
\beq
\label{eq:sigma_eps_firstcase}
\sigmahat \simeq \frac{\epsilon^2 \gamma}{4(1+\gamma)\log 2}.
\eeq

Eliminating $\epsilon$ from Equations \ref{eq:mu_eps_firstcase} and \ref{eq:sigma_eps_firstcase} reads $\mu \simeq c(\gamma, \tau)\sqrt{\sigmahat}$ in the small dissipation regime. To derive the proportionality coefficient $c(\gamma, \tau)$ we solve Eq.~\ref{eq:sigma_eps_firstcase} for  $\epsilon$ 
\beq
\label{eq:sigma_eps_firstcase2}
 \epsilon\simeq2\sqrt{\sigmahat\frac{1+\gamma}{\gamma}\log2},
\eeq
and  use Eq.~\ref{eq:mu_eps_firstcase} to find
\beq
\label{eq:cplussigmasmallA}
c(\gamma, \tau) = \sqrt{\frac{\gamma\log2}{\gamma+1}}\frac{-2e^{-\gamma\tau} + (1+\gamma)e^{-\tau}}{\gamma-1}.
\eeq

For each value of $\tau$ the function $c(\gamma,\tau)$, has a single maximum in $\gamma^*$, which is a decreasing function of $\tau$ and satisfies the transcendental equation in Eq.~\ref{eq:transsigmasmall}. In the $\gamma \rightarrow1^+$ limit, the maximum of $c(\gamma,\tau)$ reaches $\gamma^*=1$ at
\beq
\label{eq:tauC_smallsigma}
\tau_c = \frac{1+\sqrt{5}}{4},
\eeq
 (see Appendix~\ref{App:subsubsec:sigma_very_small_modelA} for details of the derivation).  For all larger values of $\tau>\tau_c$, $\gamma^*=1$ and 
\beq
c(\tau) = c(\gamma=1,\tau) =\frac{e^{-\tau}(1+2\tau)\sqrt{\log 2}}{\sqrt{2}}.
\eeq
The optimal mutual information $\mathcal{I}^*$ is linear in dissipation and exponentially decaying in $\tau$
\beq
\mathcal{I}^* \simeq \frac{c(\tau)^2 \sigmahat}{2 \log 2}.   
\eeq
The comparison with the numerical result is shown in Fig.~\ref{fig:finalAnalyticalNumericalChecks}.

\subsubsection{Limit \texorpdfstring{$\sigmahat\gg1$}{sigmahat very large}} 
\label{subsubsec:sigma_very_large_modelA}
In the large dissipation limit Eq.~\ref{eq:sigma_no-feedback} is satisfied only if $s\rightarrow 0$ with $u$ bounded by $u\leq0.5$, regardless of the initial assumptions of about $\lambda$. The optimal solution is thus in the $\gamma>1$ regime and we can extend the observations from the small $\tau$ limit to postulate that the effective magnetization is weakly dependent on the entropy production
\beqn
\mu &\simeq& c_+(\gamma,\tau)=\frac{\gamma}{\gamma^2-1}[-2e^{-\gamma\tau} + (\gamma+1)e^{-\tau}] .	\label{eq:mu_firstcase_largesigma}
\eeqn 
The effective magnetization $\mu^* \simeq c(\tau)$ has a single maximum in $\gamma$ for each $\tau$ that is a decreasing function of $\tau$ and satisfies the equation
\beq
e^{(\gamma-1)\tau} = \frac{2(1+\gamma^2 + \gamma\tau(\gamma^2-1))}{(1+\gamma)^2}.
\eeq
Similar considerations as in the small dissipation case result in
 \beq
\label{eq:tauC_largesigma}
\tau_c = \frac{1+\sqrt{3}}{2}
\eeq
above which the optimal $\gamma^*=1$. In this limit
\beq
\label{eq:mu_firstcase_largesigma_largetau}
c(\tau) = \frac{e^{-\tau}(1+2\tau)}{2}.
\eeq 
Finally, the optimal mutual information approaches a plateau for large $\sigmahat$ given by
\beq  
\mathcal{I}^* \simeq \begin{cases}
                      1+\tau\tilde{a}\log_2 \left(\tau\tilde{a}/e\right), &\gamma^*\gg 1, \tau\ll 1 \\ 
		      \frac{c(\tau)^2}{2\log2}, &\gamma^*=1, \tau\gg1
                     \end{cases}
\eeq
and is compared with the numerical result in Fig.~\ref{fig:finalAnalyticalNumericalChecks} ($\tilde{a}=0.31\mydots$ defined in Appendix \ref{App:subsubsec:sigma_very_large_modelA}).

\subsection{Feedback}
\label{subsec:diss_with_feedback}

We now ask how allowing for \textit{feedback} between the output and the input changes the energetic constraints on the optimally informative solutions. In terms of our model, this corresponds to saying that the input switching rates depend on the state of the output ($u_p\neq u_m$ and $d_p\neq d_m$), unlike in the simplest case of the model discussed in section~\ref{subsec:diss_no_feedback}. In circuits with feedback and without additional inputs the difference between the input and output is no longer clear: $z$ is an input for $x$ and vice versa.

\begin{figure}[!ht]
\hspace{- 0.1\linewidth}
\includegraphics[scale=0.7]{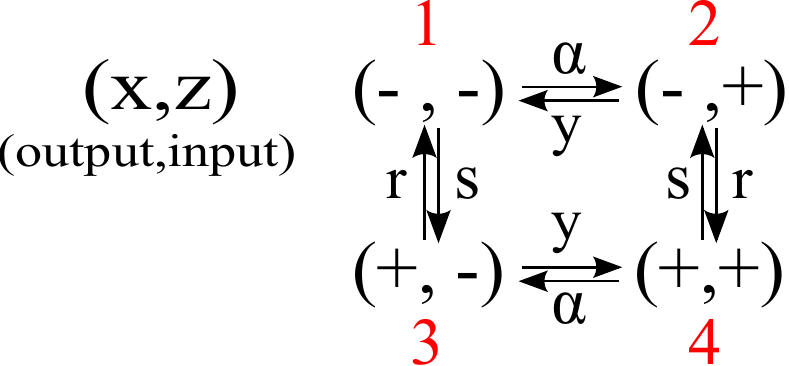}
\caption{\label{fig:drawing_modelC} The four network states, with corresponding transition rates, in a model with feedback where the input $z$ rates depend on the state of the output variable $x$.}
\end{figure}

We can exploit the symmetry between $+$ and $-$ states to decrease the number of rates in the network and set the rates of aligning (and antialigning) of the output to the input to be equal, regardless of the state of the input. Specifically, the rates defined in Fig.~\ref{fig:drawing} simplify to the ones shown in Fig.~\ref{fig:drawing_modelC}, e.~g. $r_p = r_m \equiv r = 1$, $s_p=s_m \equiv s  \leq 1$, $d_p=u_m \equiv \alpha \leq 1$, $d_m=u_m \equiv y \leq 1$. We know from our work in the infinite dissipation limit \cite{Mancini2013}, that the optimal solutions cycle irreversibly through the four states. The symmetry between $+$ and $-$ states corresponds to a degeneracy between cycling clockwise and counterclockwise and by picking this parametrization of the network we are restricting ourselves to clockwise cycles without any loss of generality.

In terms of the rates defined in Fig.~\ref{fig:drawing_modelC} the eigenvalues of $\mathcal{L}$ (see Appendix \ref{appB_diagL} for details) are $\{ \lambda_i \}=\{0,A,(A-\rho)/2,(A+\rho)/2\}$,
where
\beqn
A &=& 1+s +y+\alpha, \label{eq:A}\\
\rho &=& \sqrt{(1+s +y+\alpha)^2-8(s y+\alpha)} \label{eq:rho}.
\eeqn
The smallest nonzero eigenvalue is always $\lambda=(A-\rho)/2$ and the steady-state probability distribution is given by the normalized right eigenvector of the null eigenvalue
\beq
\label{eq:steadystate_fb}
P^{\infty} = \frac{1}{2A} \{1+y, s +\alpha, s +\alpha, 1+y\}.
\eeq
The entropy production after rescaling by the smallest eigenvalue reads
\beq
\label{eq:sigmahat_fb}
\sigmahat = \frac{2(\alpha-s y)}{A(A-\rho)}\log_2\left(\frac{\alpha}{s y}\right),
\eeq
and the mutual information is expressed by Eq.\ref{eq:MI_vs_mu} in terms of the effective magnetization 
\beqn
\label{eq:mu_fb}
&&\mu =
\exp\left(-\frac{A}{2 \lambda}\tau \right) \Big\{q \cosh\left(\frac{\rho}{2 \lambda}\tau\right)-\notag\\
&&\frac{\left[s^2-(1+y)^2-4\alpha+\alpha^2+2s(2y+\alpha)\right]}{A\rho}\sinh\left(\frac{\rho}{2 \lambda}\tau\right)\Big\}\notag,\\
\eeqn
with $q= {(1+y-s-\alpha)}/{A}$ and time rescaled by the smallest nonzero eigenvalue $\lambda$, as explained in Eq.~\ref{eq:rescMI} (see Appendix \ref{appB_mu} for  a detailed calculation of $\mu$).

The nonlinearities of the problem prohibit finding analytical solutions to this constrained optimization problem, but we explore some limiting case behavior before we turn to the full numerical optimization of the problem.

\subsubsection{Limit \texorpdfstring{$\sigma=0$}{sigma=0}}
\label{subsubsec:sigma0}

The completely equilibrium limit of $\sigmahat=0$ simplifies the constraint on the rates in Eq.~\ref{eq:sigmahat_fb} to
\beq
\alpha=s  y,
\eeq
which simplifies the effective magnetization in Eq.~\ref{eq:mu_fb} to
\beq
\mu = \frac{q}{2\rho}\left(e^{-\tau}(A+\rho) - e^{-\left(\frac{A+\rho}{A-\rho}\right)\tau}(A-\rho)\right).
\eeq
We can reparametrize the rates as
\beq
w = \frac{4 s}{(1+s )^2}, \quad v = \frac{4y}{(1+y)^2} ,  \quad  \phi = \frac{1+\sqrt{1-wv}}{1-\sqrt{1-wv}}.\quad \label{eq:w_v}
\eeq
to rewrite 
\beq
\mu = \frac{\sqrt{1-w}}{2\sqrt{1-wv}}\left(e^{-\tau} - e^{-\phi\tau}\right) + \frac{\sqrt{1-w}}{2}\left(e^{-\tau}+e^{-\phi\tau}\right).
\eeq
At $\tau=0$, the effective magnetization is $\mu=\sqrt{1-w}=\frac{1-s }{1+s }$ and is maximized for $s ^*=0$  (which implies $\alpha^*=s ^*y=0$). At $\tau>0$, the optimal effective magnetization is also obtained at $w^*=0$ (or $s^*=0$ and  $\alpha^*=0$ and $y$ is not constrained) and $\mu^*=e^{-\tau}$. The optimal mutual information $\mathcal{I}^*$ is $1$ bit for $\tau=0$ and decays in time as 
\beq
\label{eq:MI_sigma0}
\mathcal{I}^* = \frac{1}{2}\left(\log_2(1-e^{-2\tau}) + e^{-\tau}\log_2\frac{1+e^{-\tau}}{1-e^{-\tau}}\right) {\rm bits}.
\eeq

This solution corresponds to a circuit where the  two ``mixed'' states $(x,z)=\{(+,-),(-,+)\}$ are not accessible ($p_{+,-}=p_{-,+}=0$), while the two ``aligned'' states $(+,+)$ and $(-,-)$ have probability $1/2$ (see Eq.~\ref{eq:steadystate_fb}).

\begin{figure}[!ht]
\includegraphics[scale=0.6]{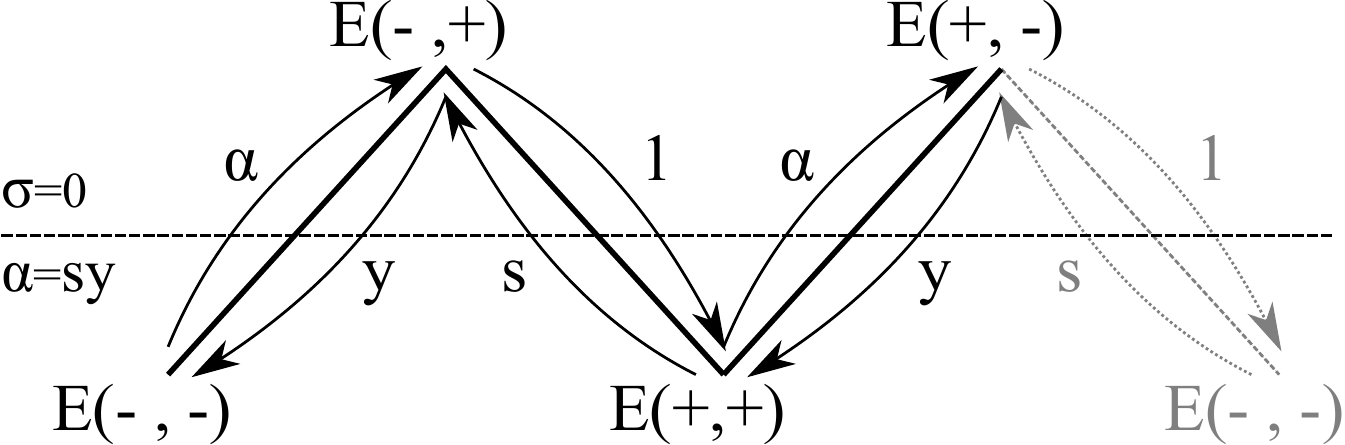}
\caption{\label{fig:drawing-sigma0} An energetic representation of the suboptimal network in perfect equilibrium for the model with feedback in Fig.~\ref{fig:drawing_modelC}. Here we show the limiting case where entropy production $\sigma=0$ and transition rates are related by the condition $\alpha=s y$, which results in $E(+,-)=E(-,+)>E(+,+)=E(-,-)$.}
\end{figure}

This optimal solution describes a completely unresponsive network with no local fluxes. The values of the nonzero rates $y$ and $r$ are irrelevant, since they account for switching from completely forbidden states. The highest possible value of information transmission is guaranteed, while remaining in an equilibrium configuration in which detailed balance is satisfied, but there is no regulation. If the readout occurs at later times, the transmitted information decays, however the nature of the solution remains the same. In summary, the optimal solution in equilibrium corresponds to a  static ``dead'' system, which is very informative since the two aligned states are on average equally sampled, but not necessary useful because the timescales for flipping between them are infinite.

The suboptimal solution at $\sigmahat=0$ differs from the optimal unresponsive network in that it has small back reaction rates for the flipping of the output and input from the antialigned to the aligned states, $s=\eta$ and $\alpha=\eta y$ (with $\eta\ll1)$. Since these reactions have non-zero rates, the ``mixed'' states have probability ${\eta}/[{(1+y)(1+\eta)}]$  and the system is able to cycle through the four states and transmit almost $1$ bit of information, without dissipating energy.

We can interpret these suboptimal solutions in terms of the energetic barriers in the system. We use the detailed-balance condition  in Eq.~\ref{eq:DB} and the Boltzmann relation, $P_i = \exp{(-{E_i}/{k_B T})}$, between the probability $P_i$ and the energy $E_i$ of state $i$ to express the rates in terms of the energies of the states and obtain the condition $E(+,-)=E(-,+)>E(+,+)=E(-,-)$, depicted in Fig.~\ref{fig:drawing-sigma0}. 

As long as the mixed states have a finite energy the system is able to cycle indefinitely through all the states  at no cost. At $s=0$ (which implies $\alpha=0$ from the equilibrium condition) infinite energy barriers separate the aligned states and lead to the unresponsive dead solution. 
When $s>0$, the input $z$ controls output $x$, transmitting information. This suboptimal costless yet informative solution is possible only because of feedback. In the simpler model of section \ref{subsec:diss_no_feedback}, the only dissipationless solution is when all rates are equal ($\alpha=y=s =r=1$ forcing $\mathcal{I}$ to be zero).

\subsubsection{Limit \texorpdfstring{$\sigmahat\ll1$}{sigmahat<<1}}
\label{subsubsec:sigma<<1}

Using the intuition from the $\sigma \equiv 0$ limit, for $\sigmahat$ nonzero but very small we expand $\alpha$ around the dissipationless solution,  as $\alpha= s  y(1-\epsilon)$ with $\epsilon\ll1$. 

For clarity, we consider the $\tau=0$ case where
\beq
\label{eq:mu_fb_tau0}
\mu = \frac{1-s +y-\alpha}{1+s +y+\alpha}.
\eeq
The rescaled dissipation, Eq.~\ref{eq:sigmahat_fb}, in terms of $\epsilon$ and the parametrizations in Eq.~\ref{eq:w_v}, is
\beq
\label{sigmaexpandedsmallsigmafeedback}
\sigmahat \simeq \frac{wv \epsilon^2}{8(1-\sqrt{wv})\log2},
\eeq
Keeping only the leading order term in $\epsilon$, the $\epsilon>0$ solution of Eq.~\ref{sigmaexpandedsmallsigmafeedback} is
\beq
\epsilon = 2\sqrt{2\sigmahat\frac{1-\sqrt{1-wv}}{wv}\log2}.
\eeq 
Expanding the effective magnetization in Eq.~\ref{eq:mu_fb_tau0} to first order in $\epsilon$ in terms of $w$ and $v$ we find
\beq
\label{mu_smallsigmafeedback}
\mu \simeq \sqrt{1-w}+ \sqrt{\sigmahat\log2}\frac{\sqrt{ w v(1-\sqrt{1-wv})}}{\sqrt{2}(1+\sqrt{1-v})}.
\eeq
The effective magnetization $\mu$ at fixed $\sigmahat$ is optimized at small but nonzero $w=\epsilon\approx 0^+$  (which translates into small but nonzero $s=\epsilon \approx 0^+$) and $v=0$ (which sets $y= 1$). Since the effective magnetization is bounded by $1$, the value of $w$ cannot be equal to zero for $\sigmahat>0$. These values set the first term in Eq.~\ref{mu_smallsigmafeedback} to $\sqrt{1-w} \approx  1$ and maximizes the coefficient of $\sqrt{\sigmahat}$, resulting in $\mu^*\approx 1$ and consequently $\mathcal{I}^*\approx 1$ bit.

Unlike in the model without feedback, it is possible to achieve almost $1$ bit of information even for arbitrarily small entropy production. Since all rates are larger than $0$, the network features all the four states, although the system spends most of its time in the aligned states $(+,+)$ and $(-,-)$. The nature of this solution is quantitatively different than the optimal unresponsive "dead" network at $\sigmahat=0$, showing that even a small amount of dissipation makes the system responsive. The optimal solution at $\sigmahat\ll1$ is also the suboptimal solution at $\sigmahat=0$.

\subsubsection{Limit \texorpdfstring{$\sigmahat\gg1$}{sigmahat>>1}}
\label{subsubsec:sigma>>1}

When entropy production is very large, we see from Eq.~\ref{eq:sigmahat_fb} that  $\sigmahat$ diverges with  $s ^*=y^*=0$. We also know from previous work \cite{Mancini2013} that mutual information is maximized for all delays $\tau$ for these values. To find $\alpha^*$, we consider the effective magnetization $\mu$ in the limit $s =y=0$:
\beq
\mu(\alpha,\tau)=\frac{1}{2(1+\alpha )\rho }\scriptstyle{\left(e^{-\tau }\frac{(1+\alpha+\rho)^2}{1-\alpha+\rho}+e^{-\tau \frac{1+\alpha +\rho }{1+\alpha -\rho }}\frac{(1+\alpha-\rho)^2}{-1+\alpha+\rho}\right)},
\eeq
where $\rho=\sqrt{1+\alpha^2-6\alpha}$.
Expanding $\mu$ to the first order in $\tau$
\beq
\mu_{\tau\ll1}(\alpha,\tau) \simeq \frac{1-\alpha}{1+\alpha}+\frac{\left(1+\alpha +\rho\right) \tau }{2(1+\alpha)}.
\eeq
we find $\alpha^*$ that maximizes the above expression
\beq
\alpha^*(\tau) = \frac{(1-\tau) \tau }{2-\tau }.
\eeq
$\alpha^*$ is an increasing function of $\tau$, until it reaches the value $\alpha^*_c=3-2\sqrt{2}$ when $\tau_c=2-\sqrt{2}$. For such value of $\alpha^*_c$, $\rho=0$ and the two smallest eigenvalues $(A-\rho)/2$ and $(A+\rho)/2$ become degenerate. Values of $\alpha$ larger than $\alpha^*_c$ are not optimal, since then $\rho$ would becomes complex and oscillations would be detrimental for information transmission \cite{Mancini2013}. 

\begin{figure*}[!ht]
\hspace{- 0.1\linewidth}
\includegraphics[scale=0.7]{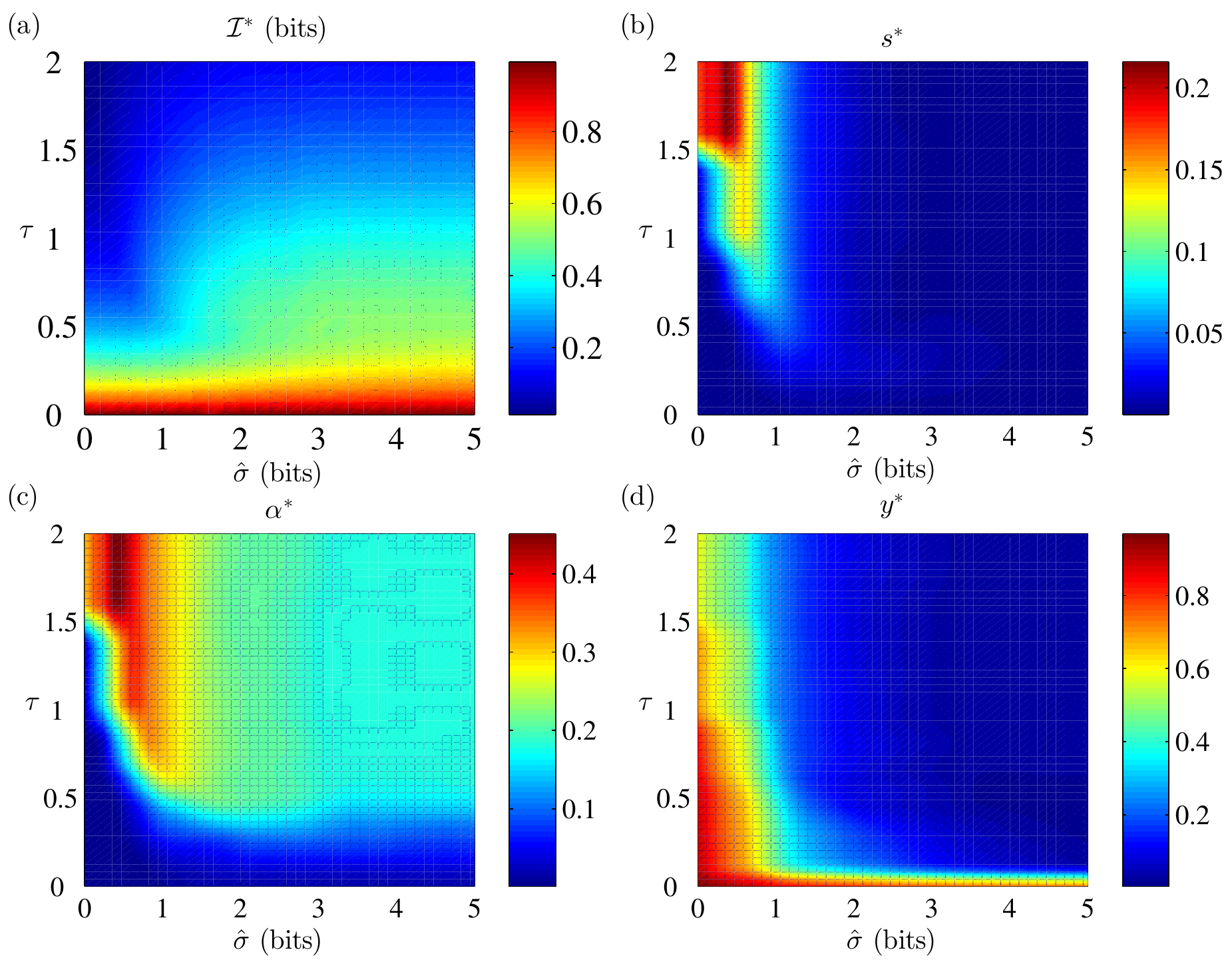}
\caption{\label{fig:contourMaxInfo_fb} 
(a) Contour plot of optimal mutual information $\mathcal{I}^*$ as function of the readout delay $\tau$ and entropy production rate $\sigmahat$, in the presence of feedback. In contrast with the simpler model of Fig.~\ref{fig:contourMaxInfo}, mutual information is now equal to $\approx$ 1 bit for any value of $\sigmahat$ when $\tau\ll1$.
(b-d) Contour plots of optimal rates $s^*$ (b), $\alpha^*$ (c) and $y^*$ (d) as functions of the readout delay $\tau$ and entropy production rate $\sigmahat$, in the presence of feedback. }
\end{figure*}

\begin{figure*}[!ht]
\includegraphics[scale=0.14]{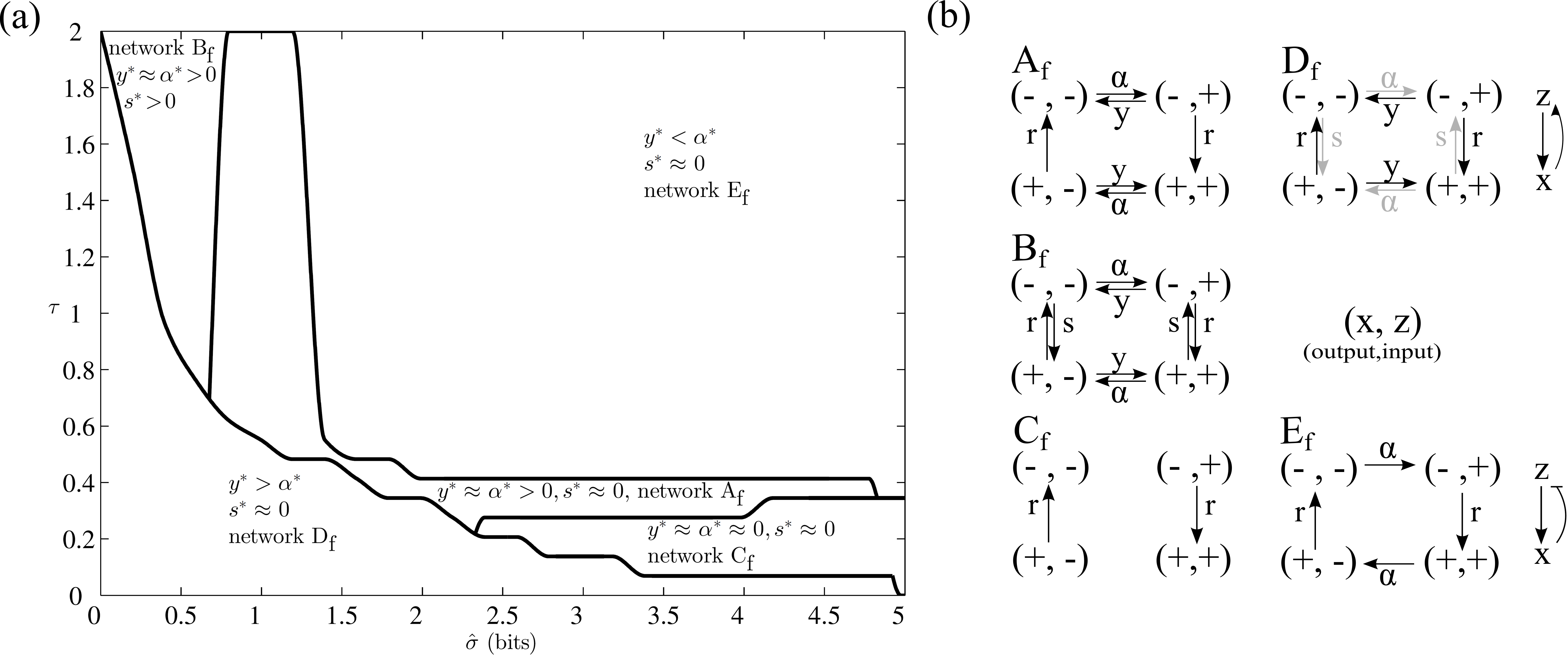}
\caption{\label{fig:phaseDiagram_modelC} \label{fig:optimalNetworks_modelC}
(a) Phase diagram in the $(\sigmahat,\tau)$ plane of optimal rates $s^*,\alpha^*,y^*$ and optimal network topologies A$_\text{f}$, B$_\text{f}$, C$_\text{f}$, D$_\text{f}$, E$_\text{f}$, in the presence of feedback. The optimal network topologies are sketched in panel (b). The gray lines in network $D_f$ of panel (b) denote the back reactions with small rates.}
\end{figure*}

\subsubsection{Numerical results}

To generalize the above results to all values of $\sigmahat$ and $\tau$ we numerically optimize the information constraining the rescaled dissipation. As in the circuits without feedback, the maximum information the circuit is able to transmit decreases with the time delay of the readout for all values of $\sigmahat$, as the system decorrelates with time  (see Fig.~\ref{fig:contourMaxInfo_fb}~a). At small but finite dissipation the decrease is exponential in $\tau$ and at large readout delays the system has similar characteristics as the circuit with no feedback:  the optimal network consists of reversible flipping of both the input and output with large rates (network $B_f$ in Fig.~\ref{fig:optimalNetworks_modelC}~b). These networks are not useful for transmitting information, but given the constraints of large time delay and close to equilibrium solution, better solutions cannot be found. As described in section~\ref{subsubsec:sigma0}, at $\sigmahat=0$ the optimal solution has the input and output 
permanently fixed in the same state, providing perfect readout but not functioning as a switch. This solution is obtained with infinitely high energy barriers between the two aligned states giving infinite switching rates between these two minima. Decreasing these energy barriers at small but finite dissipation (or for non-optimal solutions at $\sigmahat=0$), results in a finite lifetime of the two aligned states, effectively producing a stable switch with very long lived states (network $D_f$ in Fig.~\ref{fig:optimalNetworks_modelC}). These optimal networks at small but finite dissipation transmit close to $1$ bit of information, also at small but finite time delays. Feedback allows for a switching rate of the input that depends on the output and optimal circuits have fast rates for the output and input to align, and slow rates to anti-align, resulting in larger probabilities that the system is in the aligned states at the time of the readout and measurement. 
\begin{figure*}[!ht]
\hspace{- 0.1\linewidth}
\includegraphics[scale=1]{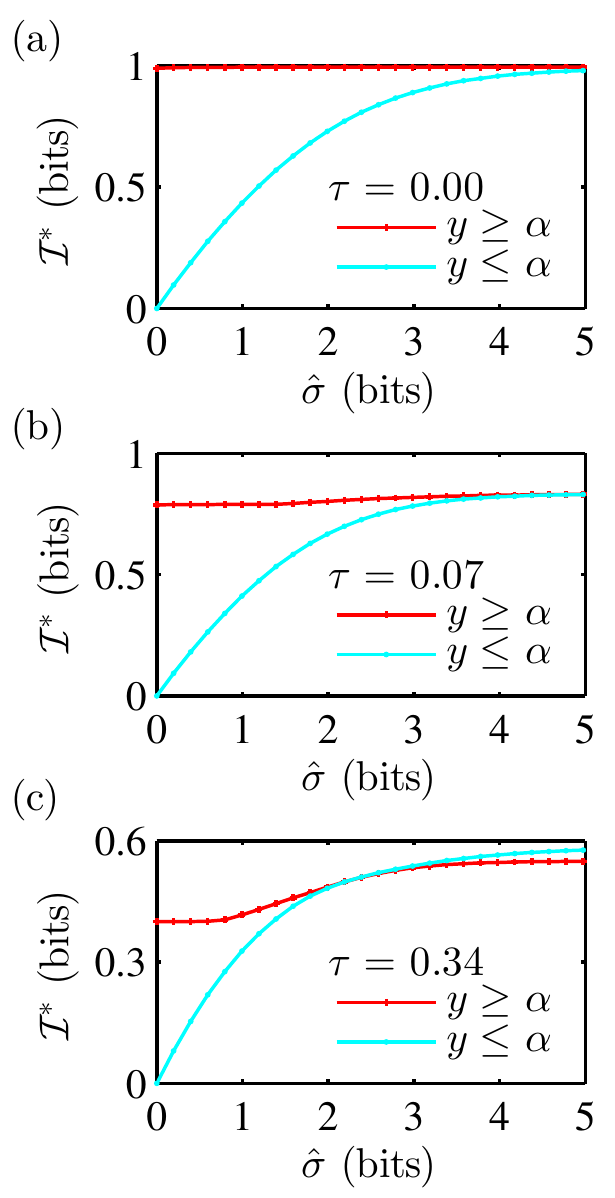}
\includegraphics[scale=1]{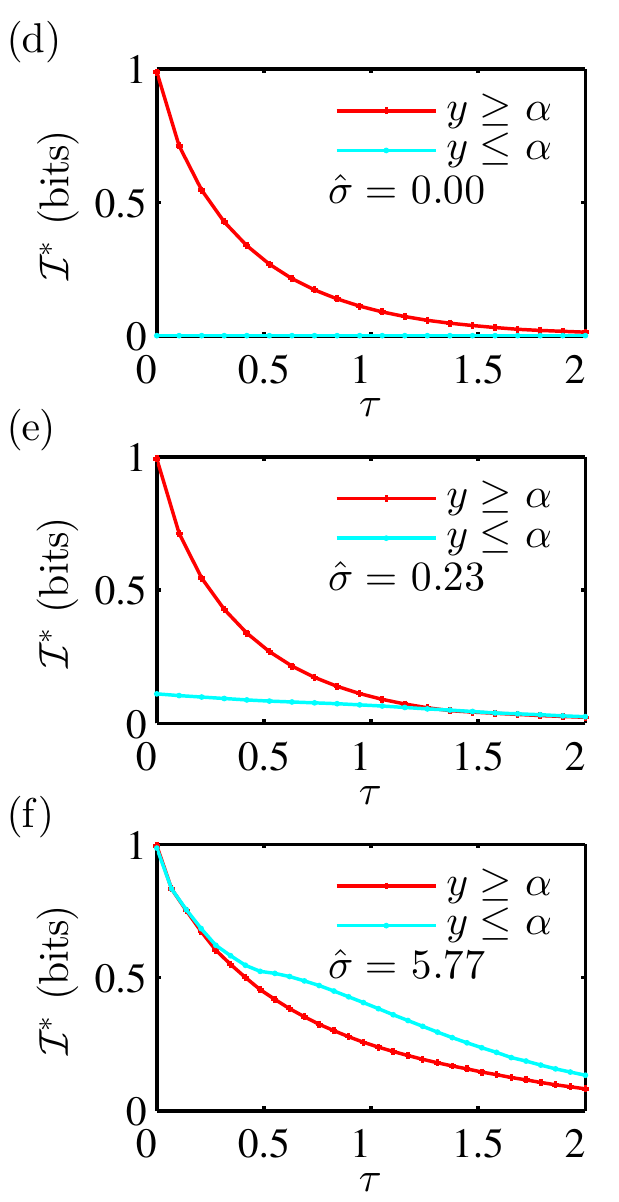}
\caption{Comparison of the mutual information for the optimal and suboptimal solutions. (a-c) \label{fig:panels_MaxInfoVsDiss} Optimal mutual information $\mathcal{I^*}$ as a function of the entropy production rate $\sigmahat$ for different readout delays $\tau$ and  (d-f) \label{fig:panels_MaxInfoVsTau}  as function of the readout delay $\tau$ for different values of the entropy production rate $\sigmahat$ (expressed in bits). Results from the simulation branch with $y\geq \alpha$ and with $y \leq \alpha$ are shown as dotted red lines and solid cyan lines, respectively. Rates used to compute such mutual information are shown in Fig.~\ref{fig:panels_RatesVsDiss} and  Fig.~\ref{fig:panels_RatesVsTau} of the section~\ref{appB}. 
The solutions of the two branches $y\geq \alpha$ and $y \leq \alpha$ coincide at large $\sigmahat$ and small  $\tau$ and at small $\sigmahat$ and large $\tau$. This happens because the back and forward input flipping rates are equal: in the first case $y^*\approx \alpha^*\approx 0$, while in the second case $y^*\approx \alpha^*> 0$ (see Fig.~\ref{fig:phaseDiagram_modelC}~a).
}
\end{figure*}
At large dissipation and small readout delays, we recover the same solution as in circuits without feedback. The input $z$ does not change and the output quickly aligns with the output (network $C_f$ in  Fig.~\ref{fig:optimalNetworks_modelC}) . As the readout time increases, the input state switches and the system decorrelates causing the transmitted information to decrease. The large dissipation rate allows the system to avoid the equilibrium solution of network $B_f$ in  Fig.~\ref{fig:optimalNetworks_modelC}, but cycle through the states with an alternating combination of fast ($r$ that aligns the input and output) and slow ($\alpha$ that anti-aligns them) rates (network $E_f$ in  Fig.~\ref{fig:optimalNetworks_modelC}). As a result the circuit is more likely to be found in the aligned states at all times, transmitting more information. As discussed above and  in our previous work \cite{Mancini2013}, the optimal network topology for large delays is a negative feedback loop, which is known to oscillate in 
certain parameter regimes \cite{Stricker2008}. Since oscillatory solutions would decrease the information transmitted at large delays, avoiding the oscillatory regime sets a limit to the maximum value of $\alpha$. As dissipation decreases in the large $\tau$ limit, the rate of aligning $z$ ($\alpha$) decreases (Fig.~\ref{fig:contourMaxInfo_fb}~c), without having a large effect on the transmitted information (network $A_f$ in Fig.~\ref{fig:optimalNetworks_modelC})~b).  Only at $\sigmahat<1$ when the rates antialigning of the input and output increase (Fig.~\ref{fig:contourMaxInfo_fb} b and c), the transmitted information decreases. 

The gain in the transmitted information per dissipation rate goes to zero at large $\sigmahat$ values, when the full non-equilibrium solution is reached, as could be expected. However also at small dissipation rates there is no increase in the transmitted information as the system dissipates more entropy. In this regime the switching rates for the input $z$ strongly favor the aligned states ($y^*>\alpha^*$), making the transition to the anti-aligned states very unlikely.  The optimal motif is a positive feedback loop. As $\sigmahat$ increases the energy barriers between the aligned and anti-aligned states decrease, since $y$ decreases, but the qualitative nature of the solution does not change. Only when the rate that favors cycling through the four states $\alpha$ increases does the transmitted information go up (and the nature of the network changes from $D_f$ to $A_f$ in Fig.~\ref{fig:optimalNetworks_modelC}~b). In this region, for intermediate values of $\tau$, the gain in transmitted information per 
increase in $\sigmahat$ is the largest.

Lastly, we compare the information transmitted by the optimal networks to that transmitted by 
suboptimal networks for characteristic values of $\tau$ and $\sigmahat$ (Fig.~\ref{fig:panels_MaxInfoVsDiss}).  
We define the suboptimal networks by dividing the optimization procedure into two branches: in one branch we constrain $y\geq\alpha$, in the other branch $y\leq \alpha$.
In this way we explore two different topologies. In the first one the system concentrates on the aligned states $(+,+)$ and $(-,-)$ (like network $D_f$ in Fig.~\ref{fig:optimalNetworks_modelC}~b), while in the second one the system cycles through the four states in the clockwise direction (like network $E_f$ in Fig.~\ref{fig:optimalNetworks_modelC}~b). From Fig.~\ref{fig:panels_MaxInfoVsDiss} we learn that for $\sigmahat \rightarrow \infty$ the optimal topology is a clockwise cycle where the system is able to transmit $1$ bit of information \cite{Mancini2013}. However, when moving towards finite values of $\sigmahat$, information transmission decreases, until a point where the system is confronted with a choice: either continue to cycle inefficiently with strong back reactions and reduce $\mathcal{I}$, or to concentrate the probability distribution on the aligned states and reach a finite plateau $\mathcal{I}=\mathcal{I}_{\sigmahat=0}$ (Fig.~\ref{fig:panels_MaxInfoVsDiss}~b). In certain cases (large 
dissipation and small and intermediate delays - Fig.~\ref{fig:panels_MaxInfoVsDiss} a and c) the two branches coincide and give the same network topologies with $s^*=0$ and small input flipping rates ($y^*\approx \alpha^*$).

\section{Robust optimization}
\label{sec:MaxiMin}

In many situations a biochemical circuit needs to reliably respond in many possible external conditions. In this case, optimization in the typical environment, as the one discussed in earlier sections, is not the desired criterium. Such a situation is better described by assuming that the environment chooses the worst possible conditions for the network to function. 
Formally this is captured by assuming that the system and the environment play a zero-sum game, where the circuit is trying to maximize the mutual information between the input and the output, while the environment is trying to minimize it. 

A game theoretic formulation of the problem requires one to define the strategy space, which in this case amounts to deciding which variables are controlled by the circuit and which by the environment control. Here we assume that the system will adjust the transition rates, whereas the environment controls the initial probability distribution $p(x_0,z_0)$ of the input $z$ and output $x$. 

In other words,  we are interested in circuits that are optimal for working in the worst possible environmental conditions, which in game theoretic terms correspond to maximin or ``minorant'' strategies \cite{Neumann1944}: the player has the goal of maximizing a function, whereas the opponent has the goal of minimizing it. This strategy is also related to ``robust control'' \cite{Chen2008,Chen2009}. In our case the circuit behaves so as to ensure that at least a certain number $\mathcal{I}$ of bits are transmitted over a given time-scale. 

\begin{figure}[!ht]
\hspace{- 0.1\linewidth}
\includegraphics[scale=0.7]{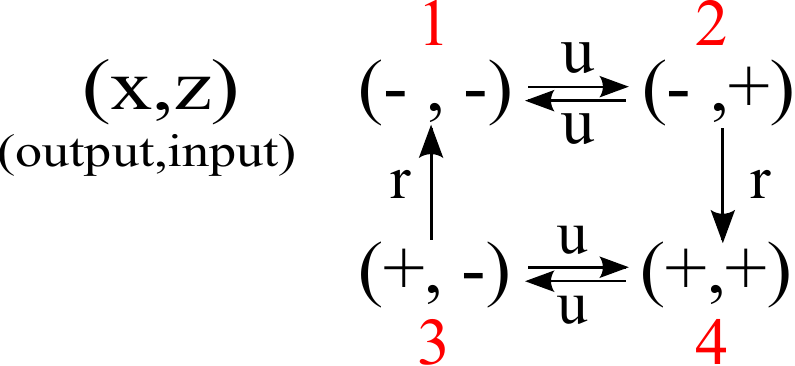}
\caption{\label{fig:drawing_MaxiMin_modelA} The four network states, with corresponding transition rates, considered in the maximin optimization where the input $z$ can either up or down-regulate the output $x$. $x$ aligns with $z$ with rate $r$.}
\end{figure}

We look for the networks that are best adapted to the worst case scenario for the simplest circuit without feedback presented in section ~\ref{subsec:diss_no_feedback} in the infinite dissipation limit. We recall that in this case the input $z$ flips between the $+$ and the $-$ state with rate $u$, and the output responds to the input with rate $r$ (see Fig.\ref{fig:drawing_MaxiMin_modelA}). Since in the infinite dissipation limit the most informative solutions always forbid the anti aligning of the input and output ($s=0$), for simplicity we consider only circuits with this constraint. 

We consider this problem on the timescales of the system, which means that the system wants to maximize the mutual information $\mathcal{I}(\tau)$ between the input at time $0$ and the output at a time $\tau=t \lambda$, where $\lambda=\min(r,2u)$ is the minimal non-zero eigenvalue of the transition rate matrix --  the inverse of the system's slowest timescale. As in section~\ref{subsec:diss_no_feedback}, we set $r=1$ to set the units of time. The effective magnetization in Eq.~\ref{eq:mumodelA} derived based on the quantities in  Appendix \ref{appC} with $s$ set to zero is:
\beq
\label{eq:MaxiMin_generalmu}
\mu=\mu_0e^{-\tau/\lambda}+\frac{1}{1-2u}\left(e^{-2u\tau/\lambda}-e^{-\tau/\lambda}\right).
\eeq
where $\mu\geq0$, and $\abs{\mu_0} \leq 1$ encodes the initial condition
\beq
\label{eq:P_0}
P(x_0,z_0)=\frac{1+x_0z_0\mu_0}{4}.
\eeq
Unlike in the cases when we optimized the transmitted information between the input and output for circuits that are in steady state in sections~\ref{subsec:diss_no_feedback} and \ref{subsec:diss_with_feedback}, in the setup considered here the initial distribution does not need to be in steady state. The space of solutions considered here is the same as the one we considered previously \cite{Mancini2013}, when we optimized the information transmitted with a delay in circuits that were out of steady state. There we simultaneously found the optimal initial distribution and the parameters of the circuit. Here, we vary the same properties of the system (initial distribution and flipping rates), but with a different underlying optimization criterium -- the environment minimizes the transmitted information by setting the initial distribution and the circuit sets the flipping rates. 

Maximizing the information transmitted in the worst case scenario in terms of this model takes the form:
\begin{itemize}
  \item The environment $E$ chooses $\mu_0$ so as to minimize mutual information, given the rates of the circuit. This corresponds to finding the value of $\mu_0$ which makes $\mu$ as small as possible (since $\mathcal{I}$ is an increasing function of $\mu$ in the allowed $\mu>1$ regime).
  \item Given $\mu_0$, the circuit $S$ looks for the rate $u$ that maximizes $\mathcal{I}$ (i.e. $\mu$).
\end{itemize}

The above zero-sum game between the system (circuit) $S$ and environment $E$  is formalized in terms of their respective cost functions $\mathcal{F}_S$ and $\mathcal{F}_E$ that satisfy 
\beq
\mathcal{F}_S + \mathcal{F}_E = 0,
\eeq
where $\mathcal{F}_S = -\mathcal{F}_E = |\mu| = \mathcal{F}(\mu_0,u;\tau)$. The optimization problem becomes
\beq
\max_{u} \min_{\mu_0} \mathcal{F}(\mu_0,u;\tau).
\eeq
The optimal $\mu_0^*$ chosen by the environment is a function of $u$ and $\tau$, such that
\beq
\min_{\mu_0} \mathcal{F}(\mu_0,u;\tau) = \mathcal{F}(\mu_0^*(\tau,u),u;\tau),
\eeq
and the circuit chooses $u^* = u^*(\tau)$ that satisfies
\beq
\max_u \mathcal{F}(\mu_0^*(\tau,u),u;\tau) = \mathcal{F}(\mu^*(\tau,u^*),u^*,\tau).
\eeq
To make analytical progress we have to separately consider the regimes of the two possible smallest eigenvalues $\lambda=\min(1,2u)$

\subsection{Case \texorpdfstring{$\lambda\leq1$}{lambda<=1}}

In the regime where $\lambda=2u\leq1$, the input switches on slower timescales then the output and the effective magnetization in Eq.~\ref{eq:MaxiMin_generalmu} is
\beq
\label{muminimaxless1}
\mu(\mu_0,u;\tau) = \mu_0 e^{-\tau/2u} + \frac{1}{1-2u}\left(e^{-\tau}-e^{-\tau/2u}\right).
\eeq
The best strategy for the environment $E$, would be to choose $\mu_0^*$ such that $\mu=0$. However it is constrained to fulfill $-1\leq\mu_0^*\leq1$. Minimizing Eq.~\ref{muminimaxless1} with respect to $u$  subject to the constraint on $\mu_0$ results in:
\beq
\mu_0^* = 
\begin{cases}
-\frac{e^{\tau(1-2u)/2u}-1}{1-2u},& \tau<\tau_c(u), \\
-1, &\tau\geq\tau_c(u),
\end{cases}
\eeq
with 
\beq
\label{eq:MaxiMin_tauc_lambdaleq1}
\tau_c(u) = \frac{4u}{1-2u}\log(1-u).
\eeq
When $\tau<\tau_c(u)$, the environment is able to set $\mu$ and thus $\mathcal{I}$ to zero. However, when $\tau\geq\tau_c(u)$, the magnetization is
\beq
\mu(-1,u;\tau) = - e^{-\tau/2u} + \frac{1}{1-2u}\left(e^{-\tau}-e^{-\tau/2u}\right),
\eeq
where $u$ is constrained to be in the interval $[0,\min(1/2,u_c(\tau))]$ and $u_c(\tau)$ is obtained by inverting Eq.~\ref{eq:MaxiMin_tauc_lambdaleq1}. Given these forms of $\mu_0^*(u,\tau)$ the circuit tries to maximize the information by tuning $u$ at each value of $\tau$. In the  $\tau\geq\tau_c(u)$ regime the effective magnetization is maximized by a $u^*$ that solves
\beq
\label{eq:condustarminimax}
\frac{\partial \mu(-1,u;\tau)}{\partial u}|_{u^*} = 0.
\eeq
Generally Eq.~\ref{eq:condustarminimax} needs to be solved numerically, but in the limit $\tau\ll1$ we find that  when $\tau\rightarrow 0$, $u^*=\frac{\tau}{2(a^*+\tau)} \rightarrow 0$  sublinearly (see section~\ref{appC_smalltauexp} for details of the derivation)
\beq
\label{uasymmaximin}
u^* \simeq \frac{\tau}{2(-\log\tau + \log(2(\log\tau)^2) -2\frac{\log(2(\log\tau)^2)}{\log\tau})}, \tau \ll1.
\eeq
The above solution $u^*$ of Eq.~\ref{eq:condustarminimax} is valid as long as the smallest eigenvalue $\lambda=2u<1$. This choice of $\lambda$ constrains $u^*<1/2$, which also constrains $\tau<\tau^*$. Setting
\beq
\frac{\partial \mu(-1,u;\tau)}{\partial u}|_{u^*=1/2} = 0, 
\eeq
we get the condition for $\tau^*=\tau(u^*=1/2)$ 
\beq
\frac{1}{2}\tau^*(\tau^*-4)e^{-\tau^*} = 0,
\eeq
which is fulfilled by $\tau^*=4$. 

In summary, the environment $E$ chooses $\mu_0^*$ so as to have $\mu(\mu_0^*,u;\tau)=0$.  However, this is possible only for $\tau<\tau_c(u)$. In this regime the transmitted information is always zero and there is nothing the circuit can do against the judicious choice of the environment. For $\tau\geq\tau_c(u)$ the best thing the environment $E$ can do is to set $\mu_0^*=-1$. In order to counteract the strategy of the environment $E$, at each readout delay $\tau$ the circuit $S$ chooses $u<u_c(\tau)$ (with $u_c(\tau)$ obtained by inverting Eq.~\ref{eq:MaxiMin_tauc_lambdaleq1}), such that the environment $E$ is forced into the regime where the best it can do is $\mu_0^*=-1$. In this regime, the circuit $S$  maximizes the function $\mu(-1,u;\tau)$ in $u\in[0,\min(1/2,u_c(\tau))]$ and finds $u^*=\frac{\tau}{2(a^*+\tau)}$, where $a^*$ is given by the solution of Eq.~\ref{eq:MaxiMin_eq_a}. The maximum value of the flipping rate for the input $u^*(\tau)=1/2$ corresponds to the readout delay $\tau^*=4$ and marks 
the transition to the regime with $\lambda=r=1$. The effective magnetization $\mu^*$ at the transition is $3/e^4\approx 0.05$ and hence $\mathcal{I}^*\approx 0.002$.

\subsection{Case \texorpdfstring{$\lambda=1$}{lambda=1}}

For $\tau>4$,  the smallest eigenvalue is $\lambda=r=1$, the input switches on faster timescales than the output and the effective magnetization in  Eq.~\ref{eq:MaxiMin_generalmu} is
\beq
\mu = \mu_0 e^{-\tau} + \frac{1}{1-2u}\left(e^{-2u\tau}-e^{-\tau}\right).
\eeq
The environment $E$ chooses $\mu_0^*$ to simultaneously set $\mu=0$ and fulfill $-1\leq\mu_0^*\leq1$, which gives
\beq
\mu_0^* = 
\begin{cases}
-\frac{1-e^{-\tau(2u-1)}}{2u-1},& \tau<\tau_c(u), \\
-1, &\tau\geq\tau_c(u),
\end{cases}
\eeq
with 
\beq
\label{eq:MaxiMin_tauc_lambda1}
\tau_c(u) = \frac{1}{2u-1}\log\left(\frac{1}{2(1-u)}\right).
\eeq
If the system $S$ wants to be in the regime $\tau\geq\tau_c(u)$ where $\mu_0^*=-1$, then the circuit must choose a rate $u^*\in[1/2,u_c(\tau)]$, with $u_c(\tau)$ obtained by inverting Eq.~\ref{eq:MaxiMin_tauc_lambda1}. This choice results in the effective magnetization
\beq
\label{muminimiaxgreat1}
\mu(-1,u;\tau) = \frac{2(1-u)e^{-\tau}-e^{-2u\tau}}{2u-1}.
\eeq
For any $\tau>4$, the effective magnetization in Eq.~\ref{muminimiaxgreat1} is always maximum at the border $u^*=1/2$. 

In summary, for $\tau>4$, the optimal response of the circuit is to set $u^*=1/2$, forcing the environment into the $\tau>\tau_c$ regime where the transmitted information is larger than zero.

\subsection{Robust Optimization Solutions}

\begin{figure}[!ht]
\includegraphics[scale=0.5]{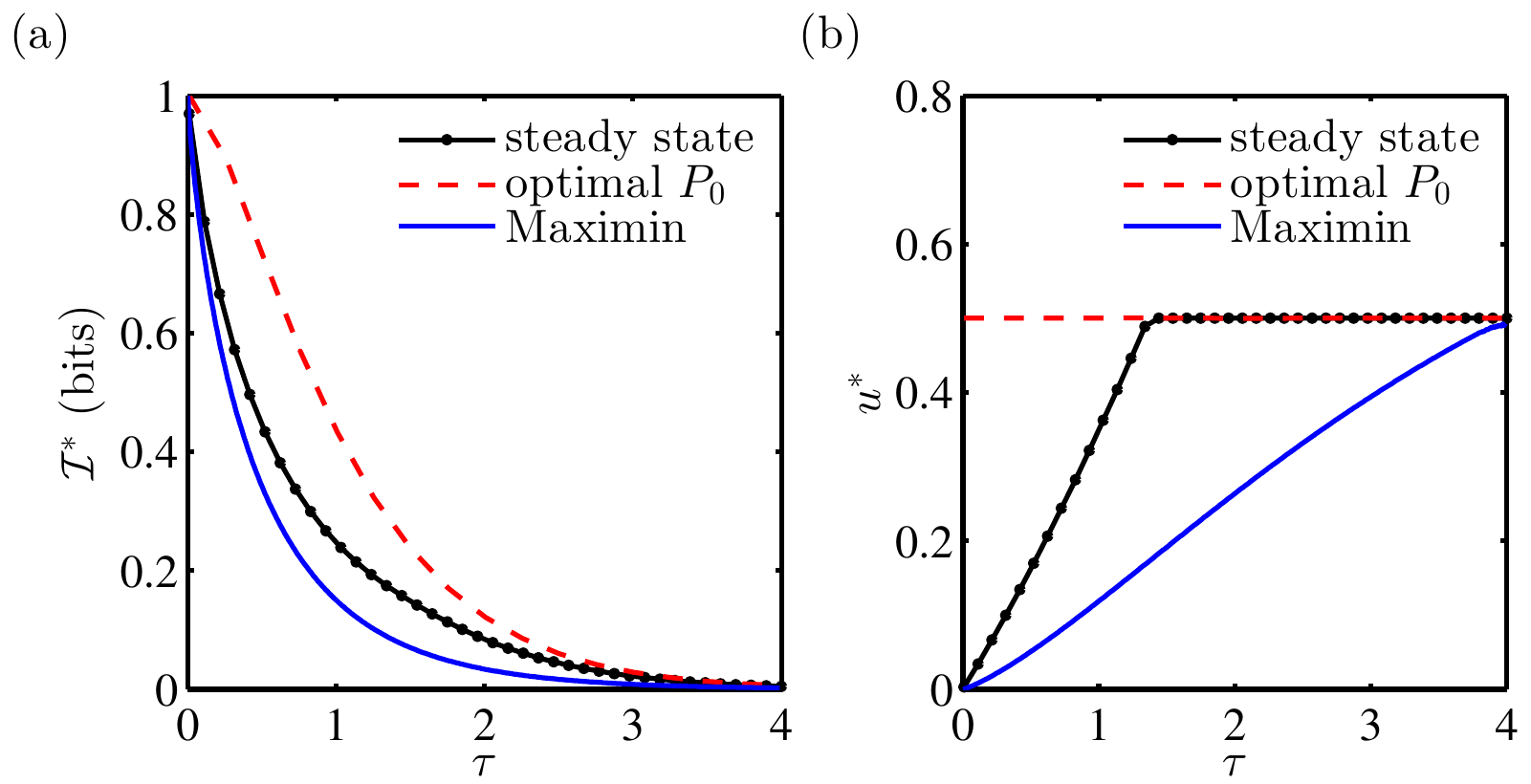}
\hspace{- 0.1\linewidth}
\includegraphics[scale=0.45]{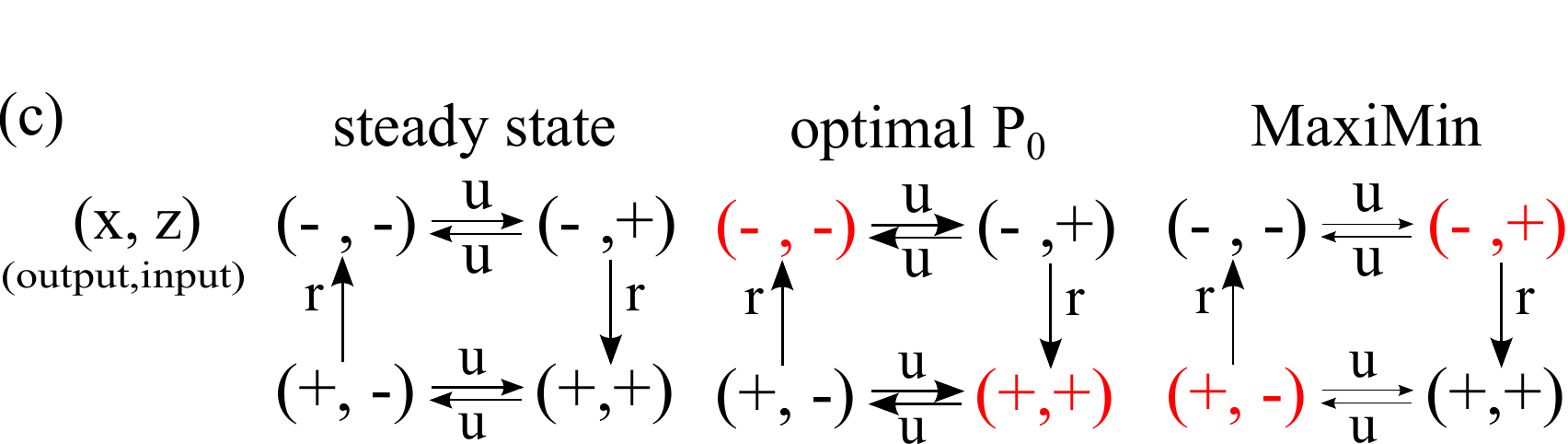}
\caption{\label{fig:MaxiMin_panels_compareImaxRates} Optimal mutual information $\mathcal{I}^*$ (a) and optimal input flipping rate $u^*$ (b) when the initial condition $P_0$ corresponds to the stationary state (dotted black line), is optimized by the system (dashed red line) or is set by an antagonistic environment in a maximin game (solid blue line). In panel (c) the optimal topology is shown in the three cases: states in red are the ones with initial probability $P_0=1/2$. Each arrow's thickness is related to the magnitude of the corresponding rate at a fixed delay $\tau=1$.}
\end{figure}

In Fig.~\ref{fig:MaxiMin_panels_compareImaxRates} we compare the capacities and optimal input switching rates  at fixed readout delay $\tau $ obtained for circuits optimized given fixed best (broken red line -- results from model $\tilde{A}$ in previous work \cite{Mancini2013}) and worst (solid blue line -- the maximin strategy discussed in this section) initial conditions to the results of simply optimizing information given the system is in steady state in the infinite dissipation regime presented in section~\ref{subsec:diss_no_feedback} (dotted black line).
In the first case the environment first fixes the initial probability distribution that is most limiting (blue line) or most favorable (red line) for information transmission and the circuit then finds the switching rates that allow it to transmit the most information, possibly neutralizing the harm of the environment. In the second case, the initial probability distribution is fixed at steady state and the circuit optimizes its switching rates within this constraint. We find that $\tau_c$ in Eq.~\ref{eq:MaxiMin_tauc_lambdaleq1} is always zero, such that the worst initial condition always corresponds to $\mu_0=-1$  for all $\tau$ and the initial probability distribution is evenly divided between the mixed states $\{(-,+), (+,-)\}$, such that $p_0(i)=1/2$. The best initial condition has $\mu_0=+1$ and the initial  probability distribution $p_0(i)=1/2$  for the aligned states $\{(+,+), (-,-)\}$. In the latter case $u^*=1/2$ for all readout delays, and the circuit functions in a regime where the input 
timescale $2u$ and the output timescale $r=1$ always match. If the initial distribution is the steady state, $u^*$ is equal to $0$ at $\tau=0$ and increases with $\tau$ until reaching the plateau $u^*=1/2$ for $\tau=({1+\sqrt{3}})/{2}$. If the environment sets the initial distribution to be the worst possible for information transmission by the circuit, $u^*=0$ for $\tau=0$ and increases much more slowly in $\tau$ than in the steady state circuit, finally converging to $u^*=1/2$ for $\tau=4$. When the circuit controls the choice of the initial state, it maximizes the probability of being in the aligned states, so that output $x$ matches input $z$ and the timescales of their switching are equal. However, when the environment chooses the worst initial state, forcing the initial probability distribution to be in the mixed states, the circuit requires the output $x$ to react as fast as possible to the input $z$ ($r\gg2u$) to align them. Despite these differences, in all cases the optimal network takes the form 
of a the same universal network (see Fig.~\ref{fig:MaxiMin_panels_compareImaxRates}~c).

\begin{figure}[!ht]
\hspace{- 0.1\linewidth}
\includegraphics[scale=0.6]{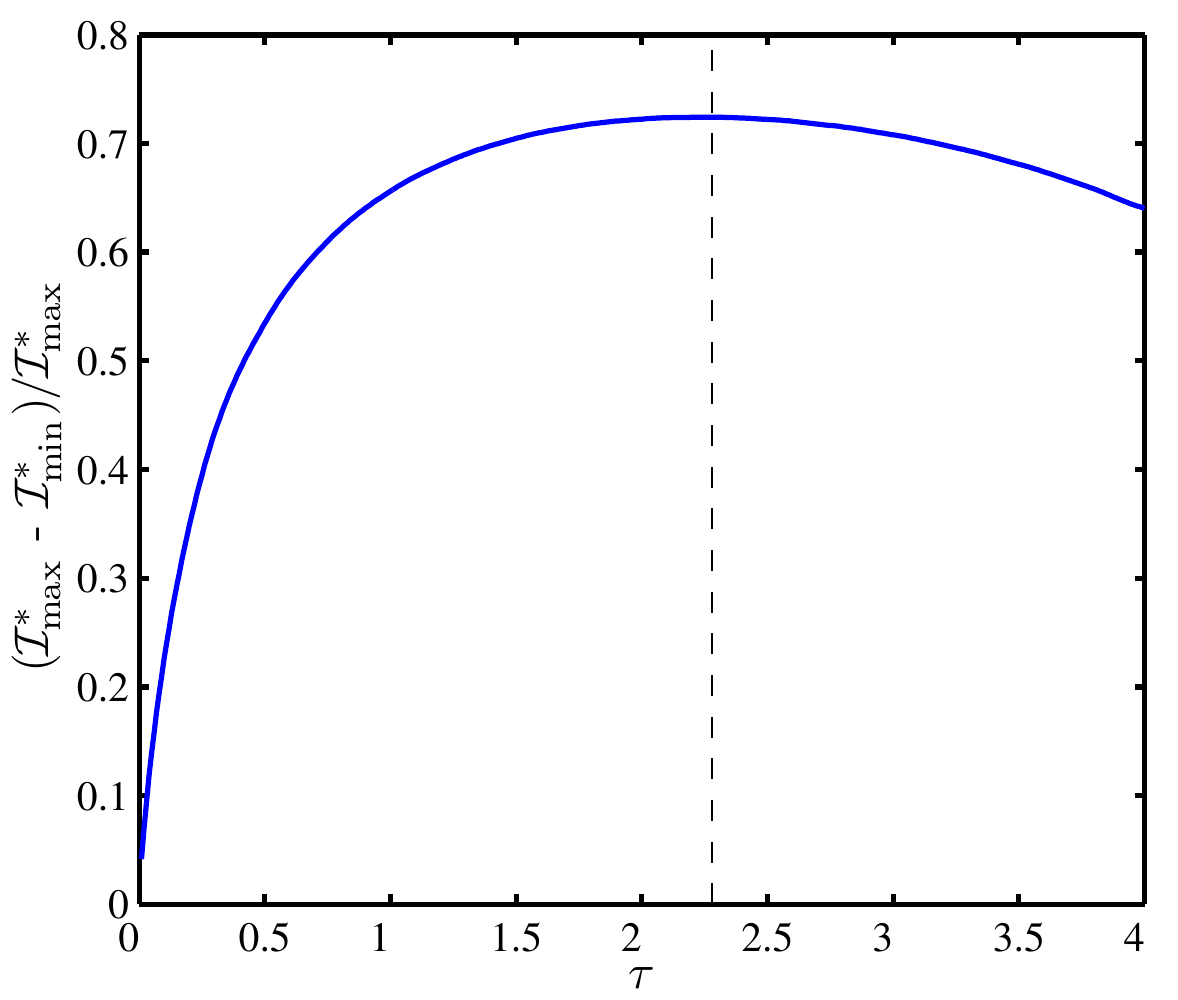}
\caption{\label{fig:MaxiMin_deltaMI} Normalized information transmission $(\mathcal{I}^*_{\max}-\mathcal{I}^*_{\min})/\mathcal{I}^*_{\max}$ as function of the readout delay $\tau^*$. Optimal $\mathcal{I}^*_{\max}$ corresponds to the case where the system optimizes the initial condition $P_0$, while $\mathcal{I}^*_{\min}$ corresponds to the MaxiMin solution, where the environment chooses the worst possible $P_0$.}
\end{figure}

The steady state $\mathcal{I}^*$  lies in between the optimal information in the maximin case ($\mu_0=-1$), which we will call $\mathcal{I}^*_{\min}$, and the one where the prior is optimized ($\mu_0=+1$), which we will indicate as $\mathcal{I}^*_{\max}$. At $\tau=0$, all three networks transmit $1$ bit of information. The maximal normalized gain $(\mathcal{I}^*_{\max}-\mathcal{I}^*_{\min})/\mathcal{I}^*_{\max}$ from optimizing the initial condition compared to the worst possible initial condition the environment can choose has a maximum at the readout delay of $\tau \approx 2.5$ (see Fig.~\ref{fig:MaxiMin_deltaMI}). At this timescale the environment can be most detrimental for information transmission.

\section{Discussion}
\label{sec:discussion}

Most studies that optimize information transmission in biochemical circuits consider ideal conditions and look for the networks that are only limited by intrinsic physical constraints coming from noise in the system. However often cells must respond to signals under natural external constraints: the readout of the input occurs at a delay, cell energetics are limited and the environment may be unfavorable -- it need not be tuned to the properties of the network. Here we investigated how these difficulties influence the form of optimal designs of biochemical circuits. 

Most generally, the information transmitted by circuits decreases with the readout delay, as the system decorrelates with time. Feedback can decrease this decays, but cannot overcome it completely. In the large dissipation limit the optimal solution consist of using a combination of fast rates for output switching and slow rates for input switching to increase the probability of the system to be in two states. Our choice of setting the input rate 
that aligns the input and output states ($r=1$) fixed these two states to be the aligned states, but the natural symmetry of the system implies that a degenerate solution that transmits the same amount of information exists for the case when the input represses the output, favoring the anti-aligned states. We explicitly discussed these solution in the infinite dissipation regime in previous work \cite{Mancini2013}. In the simplest circuit without feedback the only way to achieve this separation into favorable and unfavorable states is by dissipating energy and forbidding back reactions for output switching. Close to equilibrium, in the absence of feedback, the circuit cannot constrain  the back reactions and as a result, the maximum mutual information goes linearly to zero with the entropy production rate $\sigmahat$ for all values of the  readout delay $\tau$. From the simplest circuit we see that the rate of input flipping depends on the time delay -- longer readouts require slow flipping rates of the 
input to be informative, whereas the ability to dissipate energy allows the circuit to irreversible cycle through the states by eliminating both the input and output back reactions. The fully non equilibrium solution is valid for a large range of dissipation values. If long readouts or energy constraints forbid this solution the circuit effectively becomes randomly stuck in one of two states and not informative: the input is fixed with an equal probability to be in one of the two states, and the output attempts to align with the input. In summary, the only way for a system without feedback to transmit information is to dissipate energy. 

Feedback significantly increases the range of dissipation values at which circuits can be informative. When the output feedbacks onto the input, the circuit can transmit $\approx 1$ bit of information for any value of $\sigmahat$ even at small delays. Far from equilibrium, the optimal solution cycles through all the states, effectively increasing the decorrelation time of the system. The optimal topology is based on a negative feedback motif with a slow switching input and rapidly responding output. Such motifs are very common in stress responses (DNA damage, heat and osmotic shock and immune response) \cite{Lahav2004} and often rely on a slow (gene regulation) and fast (protein-protein interaction) step.  In perfect equilibrium, the formally optimal circuit is non responsive -- there is no regulation and the input and output are aligned at all times. However at small but finite entropy production rates (as well as the suboptimal solution in perfect equilibrium) the optimal topologies are different from the 
large dissipation case.

In the presence of feedback, the optimal circuit in the small dissipation range is a  positive feedback loop with two stable states $(+,+)$ and $(-,-)$. Such circuits have long been known to be a key mechanism for memory storage \cite{Lisman1985}. This design of a stable switch is able to convert a transient stimulus into a permanent biochemical response.  These circuits have been shown to be crucial for the irreversibility of maturation of \textit{Xenopus} oocytes \cite{Xiong2003} and for long lasting synaptic plasticity \cite{Tanaka2008}. It has also been argued that positive feedback may have a role in enhancing switch-like responses (e.~g. in MAP kinase cascades) and improving energetic efficiency by filtering out noise \cite{Cinquin2002}. This may explain why we find such optimal topology in the small dissipation regime. 

The above examples show that the optimal topology at small dissipation rates is characteristic of stable long term readouts, that commit the cell to one of two responses.  The aligned (or anti-aligned in the other degenerate topology) states are very stable and large energetic barriers exist to exit these states, resulting in the positive feedback motif being optimal. Conversely, the optimal motif in the large dissipation limit is a negative feedback loop, that is characteristic of shock response -- a transient response that is easily exited, but needs to be implemented quickly. It is therefore a typical non--equilibrium response, whereas the positive feedback loop is characteristic of slow and stable equilibrium situations. 

Intuitively, dissipating more energy allows for larger information transmission because it lowers the probability of back reactions, which are detrimental when processing a signal. 
Interestingly, in the presence of feedback the system is able to build a particular topology which is suboptimal in terms of information transmission but which does not dissipate energy at all. 
The resulting network is such that effectively the system can cycle either in the clockwise or counterclockwise direction and the probability distribution is mostly concentrated on the aligned states $(+,+)$ and $(-,-)$. Such costless network topologies could be of inspiration when designing synthetic biochemical circuits aimed at energy production.

Feedback is able to slow down the decrease of information transmission with readout delay, but not change the monotonic nature of this process  caused by decorrelation of the states of the circuit. Yet feedback does alter the dependence of the information decay with dissipation compared to circuits without feedback. At large as well as small dissipation rates the capacity plateaus, leaving a small range of $\sigmahat$ values where the transmitted information is sensitive to the precise magnitude of the energy constraints. This relatively narrow regime is where the optimal motif changes from a positive feedback loop  to a negative feedback loop. Effectively in this regime the feedback is turned off (the back and forth input flipping rates are similar) and the circuit resembles the simple system discussed in section~\ref{subsec:diss_no_feedback}.

Our optimal network for information transmission with large energy dissipation at relatively large readout delays (circuit $E_f$ in Fig.~\ref{fig:contourMaxInfo_fb}) has the same design as the two-component signaling network in \textit{Escherichia coli} used in osmoregulation \cite{Stock1995, Egger1997,Stock2000, Barbieri2013}. This network is composed of the histidine kinase EnvZ and the response regulator OmpR and it is aimed at reacting to an osmotic shock by regulating the expression of two porin proteins OmpF and OmpC. After phosphorylation by EnvZ, OmpR undergoes a conformational change, dimerizes and binds to the porin promoter region either of the \textit{ompF} or \textit{ompC} gene.  
We can map the activation of our input $z$ to the process of phosphorylation and conformational change of OmpR and the activation of our output $x$ to the dimerization and binding to DNA of OmpR. Conversely, the deactivation of $x$  corresponds to unbinding of OmpR from the DNA, while the deactivation of $z$ to the dephosphorylation of OmpR. Detailed experimental studies of the energetics of this system show that phosphorylation-activated dimerization drives an increase in DNA binding \cite{Barbieri2013}, suggesting that the biochemical regulation is a clockwise cycle such as presented in network $E_f$ of Fig.~\ref{fig:contourMaxInfo_fb}.

In the large dissipation limit we compared three different conditions in which a circuit optimizes the information transmitted at a delay for a model without feedback: a circuit that functions in steady state (section~\ref{subsec:diss_no_feedback}), one that is able to optimize its input distribution (derived previously \cite{Mancini2013}), and one that is forced to function with the least informative initial distribution (maximin - section~\ref{sec:MaxiMin}). Interestingly, all solutions share the same circuit topology and type of solution. The most informative solution is to cycle irreversibly through the four states. The difference between the three cases lies in the rate of flipping the input signal at a given delay. The most informative of the three strategies, where the circuit has coevolved to match the environmental conditions, displays the largest flipping rate of the input (although still small compared to the flipping rate of the output)  that is independent of the readout delay. The least 
informative circuit, the one that functions in an adverse environment has the slowest flipping rate of the input. Intuitively, if the statistics of the environment and circuit match, then as long as these initial states are long lived the ability of the system to transmit information is mainly encoded in these states. However, in an adverse environment, extremely small flipping rates of the input stabilize the initial input states, allowing for a more informative readout. Since the same circuit, just with different flipping rates of the input, works optimally in both favorable and antagonistic environmental conditions, one could imagine that the rate of input switching could be tuned depending on the environmental conditions. This tuning could be achieved by fast degradation of a ''typical" sugar source (like glucose) but a slower degradation that requires additional elements (such as production of the enzyme beta-galactosidase) for degradation of a less typical sugar source (like lactose). 

The models of biochemical regulation we consider assume the limit of very sharp response functions, that simplify their description to two state systems. As was previously shown, on one hand smooth regulatory functions can transmit more than 1 bit of information \cite{Tkacik2008}, and on the other hand the molecular noise coming from discrete particle numbers limits the capacity \cite{Tkacik2009, Walczak2010, Tkacik2011, Tkacik2012, Tostevin2009, Tostevin2010, Govern2014}. The capacity and regulatory details of the optimal systems can change if we consider more detailed molecular models. However even these simple models show general principles of how energy constraints and delayed readout drive optimal topologies. It has previously been argued using more detailed models that a truly bistable system in equilibrium is not optimal for transmitting information, unless the system does not have time to equilibrate and manages to retain memory of the initial condition \cite{Tkacik2012}. The solutions we observe in 
our optimal networks with feedback at small dissipation correspond to circuits that manage to retain the memory of the initial state. 

All the models we considered, both in equilibrium and out of equilibrium, corresponded to two component systems. These types of networks were previously studied as circuits that can function out of equilibrium in contrast with one component signaling systems that must obey detailed balanced \cite{Govern2014, Govern2014a}. When it comes to precision of a continuous gradient readout, it was shown that fueling energy into the system makes it possible to overcome the limitations posed by detailed balanced, by decoupling the output and receptor molecules and providing a stable readout of the input. In our discrete two component system, this stable readout of the input state is possible even at equilibrium with a circuit design that is able to stably store the input state by exploiting timescale separation and favoring the aligned states over the non-aligned ones. However, such a stable solution is not very useful for responding to signals that change on fast timescales.  In that case, energy dissipation is 
indispensable for an informative readout.


{\bf Acknowledgements.} We thank A.~C. Barato and T.~Mora for helpful discussions. AMW is funded by a Marie Curie Career Integration Grant.

\appendix
\section{General form of the rate matrix \texorpdfstring{$\mathcal{L}$}{L}}
\label{app0}

The transition rate matrix $\mathcal{L}$ of Equation~\ref{eqn:master_eq} in the main text is given in its general form by
\beq
\label{eq:general_L}
\mathcal{L} = \left( \begin{array}{cccc}
                      u_m+s_m & -d_m & -r_m & 0 \\
		      -u_m & d_m+r_p & 0 & -s_p \\
		      -s_m & 0 & u_p+r_m & -d_p \\
		      0 & -r_p & -u_p & d_p+s_p
                     \end{array}
	      \right).
\eeq
When $\mathcal{L}$ is analytically diagonalizable, its eigenvalues $\lambda_{\alpha}$, its right eigenvectors $v_{\alpha}$ and its left eigenvectors $u_{\alpha}^T$ (with $\alpha=1,\dots,4$) satisfy
\beqn
\mathcal{L} v_{\alpha} &=& \lambda_{\alpha}  v_{\alpha} \\
u_{\alpha}^T \mathcal{L} &=& u_{\alpha}^T  \lambda_{\alpha} \\
u_{\alpha}^T v_{\beta} &=& \delta_{\alpha \beta}.
\eeqn
The steady-state probability vector $P_{\infty}$ is equal to the right eigenvector $v_1$, which corresponds to the null eigenvalue $\lambda_1$ and which is given by
\beq
\label{eq:general_ss}
v_1=\frac{1}{\sum_{i=1}^{4} v_1(i)}\left(\begin{array}{c}
\frac{d_p r_m r_p+d_m (d_p r_m+s_p (r_m+u_p))}{r_m r_p u_m+(d_m s_m+r_p (s_m+u_m)) u_p} \\ 
\frac{d_p r_m u_m+s_p (r_m u_m+(s_m+u_m) u_p)}{r_m r_p u_m+(d_m s_m+r_p (s_m+u_m)) u_p} \\
\frac{d_m s_m (d_p+s_p)+d_p r_p (s_m+u_m)}{r_m r_p u_m+(d_m s_m+r_p (s_m+u_m)) u_p}\\
1\\
\end{array}\right).
\eeq

\section{Computation of the entropy production rate \texorpdfstring{$\sigma$}{sigma}}
\label{app1}

Here we perform the calculation of the entropy production rate $\sigma$, for the general dynamic system described by the transition rate matrix $\mathcal{L}$ and pictured in Fig.~\ref{fig:drawing}.\\
We start from the definition of $\sigma$, introduced in Eq.~\ref{eq:sigma} in the main text:
\beq
\sigma= \sum_{i,j}P_i w_{ij}\log\frac{w_{ij}}{w_{ji}},
\eeq
where $P_i$ is the steady state probability distribution $P_{\infty}$ for state $i$.
In our specific case, we explicitly have
\beqn
\sigma &=& P_1 w_{12}\log\frac{w_{12}}{w_{21}} + P_2 w_{24}\log\frac{w_{24}}{w_{42}} +\\
       &+& P_4 w_{43}\log\frac{w_{43}}{w_{34}} + P_3 w_{31}\log\frac{w_{31}}{w_{13}} +\\
       &+& P_2 w_{21}\log\frac{w_{21}}{w_{12}} + P_4 w_{42}\log\frac{w_{42}}{w_{24}} +\\
       &+& P_3 w_{34}\log\frac{w_{34}}{w_{43}} + P_1 w_{13}\log\frac{w_{13}}{w_{31}}.
\eeqn
After collecting similar terms we can write
\beqn
\sigma &=& J_{12}\log \frac{w_{12}}{w_{21}} + J_{24}\log \frac{w_{24}}{w_{42}} + \\
       &+& J_{43}\log \frac{w_{43}}{w_{34}} + J_{31}\log \frac{w_{31}}{w_{13}},
\eeqn
where we have used the definition of probability current $J_{ij}$, introduced in Section \ref{sec:diss-model} in the main text, and we have considered the clockwise cycle in Fig.~\ref{fig:drawing}.
All the steady state currents are equal to each other:
\beq
J_{12} = J_{24} = J_{43} = J_{31} = J
\eeq
and the entropy production rate $\sigma$ is
\beq
\label{app1_sigma}
\sigma =  J \log\frac{w_{12}\, w_{24} \, w_{43} \, w_{31}}{w_{21}\, w_{13} \, w_{34} \, w_{42}}.
\eeq
We plug  the rates of the transition matrix $\mathcal{L}$ (see Eq.~\ref{eq:general_L}) and the stationary distribution $P_{\infty}$ (see Appendix \ref{app0}) into Eq.~\ref{app1_sigma}. The current $J$ is
\beq
J = \frac{(u_m r_m d_p r_m)-(d_m s_m u_p s_p)}{J_a+J_b+J_c},
\eeq
with
\beqn
J_a &=& d_p (r_m (r_p+u_m)+r_p (s_m+u_m)),\notag\\
J_b &=& d_m (d_p (r_m+s_m)+r_m s_p+s_m s_p+s_m u_p+s_p u_p),\notag\\
J_c &=& (r_p+s_p) (r_m u_m+(s_m+u_m) u_p),\notag
\eeqn
and entropy production rate is
\beq
\sigma =  J\log\frac{u_m r_m d_p r_m}{d_m s_m u_p s_p}.
\eeq

\section{Information transmission with energy dissipation: the simplest model}
\label{appA}

\subsection{Diagonalization of \texorpdfstring{$\mathcal{L}$}{L}}
\label{appA_diagL}
In the simplest model described in section~\ref{subsec:diss_no_feedback}, the transition rate matrix $\mathcal{L}$ has the form
\beq
\mathcal{L} = \left( \begin{array}{cccc}
                      u+s & -u & -1 & 0 \\
		      -u & u+1 & 0 & -s \\
		      -s & 0 & u+1 & -u \\
		      0 & -1 & -u & u+s
                     \end{array}
	      \right).
\eeq
Its eigenvalues $\lambda_{\alpha}$ are 
\beq
\begin{cases}
\lambda_1 = 0 \\
\lambda_2 = 2u \\
\lambda_3 = 1+s \\
\lambda_4 = 2u+1+s
\end{cases},
\eeq
its right eigenvectors $v_{\alpha}$ are
\beqn
P_{\infty}=v_1 &=&\frac{1}{2(1+s+2u)}\left(\begin{array}{c} u+1 \\ u+s  \\ u+s  \\ u+1  \end{array}\right),\\
v_2 &=& \frac{1}{2(1+s-2u)}\left(\begin{array}{c} u-s \\ u-s  \\ s-u  \\ s-u  \end{array}\right),\notag\\
v_3 &=&\frac{1}{2(1+s-2u)}\left(\begin{array}{c} u-1 \\ s-u  \\ s-u  \\ 1-u  \end{array}\right),\\
v_4 &=& \frac{u+s}{2(1+s+2u)}\left(\begin{array}{c} +1 \\ -1  \\ -1  \\ +1  \end{array}\right).
\eeqn
and its left eigenvectors $u_{\alpha}^T$ are
\beq
\begin{cases}
u_1^{T} = (1,1,1,1), \\
u_2^{T}= (-1,\frac{-u+1}{+u-s},\frac{+u-1}{+u-s},1), \\
u_3^{T} = (-1,1,-1,1),\\
u_4^{T} = (1,\frac{-u-1}{+u+s},\frac{-u-1}{+u+s},1).
\end{cases}
\eeq
The stationary state $P_{\infty}$ is equal to the first right eigenvector $v_1$, related to the null eigenvalue $\lambda_1$.

\subsection{Limit \texorpdfstring{$\tau\ll1$}{tau very small}}
\label{appA_tausmall}

Here we detail the solutions of the equations presented in the small $\tau$ limit of the simplest model in section \ref{subsec:diss_no_feedback}. In the small dissipation limit we assume 
\beq
\label{appA_gammastar}
\gamma^* \simeq \frac{a_0}{\tau} + b_0 + c_0 \tau
\eeq
 and find the coefficients of the expansion by solving $\frac{d\mu}{d\gamma}|_{\gamma^*}=0$ (with $\mu$ given by Eq.~\ref{eq:mu_smalltau_smallsigma}) order by order in $\tau$. The coefficient $a_0=0.96\mydots$ is given by the solution of the transcendental equation $e^{a_0}=\frac{4}{3}(a_0+1)$, while $b_0$ and $c_0$ are given respectively by
\beqn
b_0 &=& 1 + \frac{-5-2a_0}{6a_0^2} = -0.25\mydots \quad \text{and}\notag\\
c_0 &=& \displaystyle{\frac{-75-65 a_0 + 128 a_0^2 + 28 a_0^3}{72 a_0^5}} = 0.1\mydots\notag
\eeqn
Using the Eq.~\ref{appA_gammastar} for $\gamma^*$ gives
\beq
\mu^* \simeq \sqrt{\sigmahat}(1 + A_0 \tau + B_0 \tau^2),
\eeq
with
\beqn
A_0 &=& \frac{-4e^{-a_0} + 3}{2a_0}-1 = -0.24\mydots \quad \text{and}\notag\\
B_0 &=& \textstyle{e^{-a_0}\frac{-5-7a_0+a_0^2+6a_0^3}{3a_0^4} + \frac{10+4a_0-a_0^2-12a_0^3+4a_0^4}{8a_0^4}}= 0.01\mydots\notag
\eeqn

Similarly in the large dissipation limit we take $\gamma^*\simeq \frac{a_{\infty}}{\tau} + b_{\infty} + c_{\infty} \tau$ and following the same procedure as above with $\mu$ given by Eq.~\ref{eq:mu_smalltau_largesigma} we find $a_{\infty}= 1.68\mydots$ as the solution of the transcendental equation $e^{a_{\infty}}=2(a_{\infty}+1)$, while $b_{\infty}$ and $c_{\infty}$ are given by
\beqn
b_{\infty} &=& 1 + \frac{-2-a_{\infty}}{a_{\infty}^2}= -0.31\mydots \quad \text{and}\notag\\
c_{\infty} &=& \displaystyle{\frac{-12-12 a_{\infty} + 7 a_{\infty}^2 + 3 a_{\infty}^3}{2 a_{\infty}^5}}= 0.07\mydots\notag
\eeqn
The effective magnetization is
\beq
\mu^* \simeq 1 + A_{\infty} \tau + B_{\infty} \tau^2,
\eeq
with
\beqn
A_{\infty} &=& \frac{-2e^{-a_{\infty}} + 1}{a_{\infty}}-1 = -0.63\mydots \quad \text{and} \notag\\
B_{\infty} &=& \textstyle{\frac{1}{2} + \frac{(e^{-a_{\infty}}(-4-6a_{\infty}+2a_{\infty}^2) + 2+a_{\infty}-a_{\infty}^3)}{a_{\infty}^4}}= 0.23\mydots\notag
\eeqn

\subsection{Limit \texorpdfstring{$\sigmahat\ll1$}{sigmahat very small}}
\label{App:subsubsec:sigma_very_small_modelA}

In the limit of $\sigmahat\ll1$ we assume that $\mu \simeq c(\gamma, \tau)\sqrt{\sigmahat}$ and $s=1-\epsilon$, generalizing the $\sigmahat=0$ behavior. Solving Eq.~\ref{eq:sigma_eps_firstcase} for $\epsilon$ we obtain the form of $c(\gamma,\tau)$ in Eq.~\ref{eq:cplussigmasmallA}. For each value of $\tau$ the function $c(\gamma,\tau)$, has a single maximum in $\gamma^*$, which is a decreasing function of $\tau$ and satisfies the transcendental equation
\beq
\label{eq:transsigmasmall}
e^{(\gamma^*-1)\tau} = 2\frac{1+\gamma^*+2\gamma^{*2} + 2\gamma^*\tau(\gamma^{*2}-1)}{(1+\gamma^*)(1+3\gamma^*)}.
\eeq
The maximum of $c(\gamma,\tau)$ is found from $d c/d\gamma \sim F(\gamma, \tau)/(\gamma-1)^2=0$, where $F(\gamma, \tau)=0$ is solved by Eq.~\ref{eq:transsigmasmall}, for $\gamma >1$. In the limit $\gamma \rightarrow1^+$, the maximum is found as the solution of $e^{-\tau}\left(1+2 \tau -4 \tau ^2\right)/(4 \sqrt{2})=0$, and at 
\beq
\tau_c = \frac{1+\sqrt{5}}{4},
\eeq
 the maximum of $c(\gamma,\tau)$ reaches $\gamma^*=1$. 

\subsection{Limit \texorpdfstring{$\sigmahat\gg1$}{sigmahat very large}}
\label{App:subsubsec:sigma_very_large_modelA}

In the limit of large dissipation $\sigmahat$ and delay $\tau\ll1$ it is possible to write the optimal mutual information as
\beq  
\mathcal{I}^* \simeq 1+\tau\tilde{a}\log_2 \left(\tau\tilde{a}/e\right)
\eeq
where $\tilde{a}=0.31\mydots$ is defined as
\beq
\tilde{a} = \frac{a_{\infty}}{2(a_{\infty}+1)},
\eeq
with $a_{\infty}=1.68\mydots$ introduced in Appendix \ref{appA_tausmall}.

\section{Information transmission with energy dissipation: feedback}
\label{appB}

\subsection{Diagonalization of \texorpdfstring{$\mathcal{L}$}{L}}
\label{appB_diagL}

In the model where feedback is present (see section \ref{subsec:diss_with_feedback}), the transition rate matrix $\mathcal{L}$ has the form
\beq
\mathcal{L} = \left(
\begin{array}{cccc}
 s+\alpha  & -y & -1 & 0 \\
 -\alpha  & 1+y & 0 & -s \\
 -s & 0 & 1+y & -\alpha  \\
 0 & -1 & -y & s+\alpha 
\end{array}
\right).
\eeq\\
It is useful to introduce the quantities $A$ and $\rho$, defined in the main text as
\beqn
A &=& 1+s+y+\alpha \label{eq:appB_A}\\
\rho &=& \sqrt{(1+s+y+\alpha)^2-8(s y+\alpha)} \label{eq:appB_rho}.
\eeqn
Then the eigenvalues $\lambda_{\alpha}$ can be written as
\beq
\begin{cases}
\lambda_1 = 0 \\
\lambda_2 =  A \\
\lambda_3 =  \frac{1}{2}(A-\rho) \\
\lambda_4 = \frac{1}{2}(A+\rho)
\end{cases}.
\eeq
The right eigenvectors $v_{\alpha}$ are given by
\beqn
P_{\infty}=v_1 &=& \frac{1}{2A}\left(\begin{array}{c} 1+y \\ s+\alpha \\ s+\alpha \\ 1+y \end{array}\right),\label{eq:appB_v1}\\
v_2 &=& \frac{s+\alpha}{2A} \left(\begin{array}{c} +1 \\ -1 \\ -1\\ +1 \end{array}\right),\\
v_3 &=& \frac{1}{4\rho}
				\left(\begin{array}{c} 
					+(-1+s-y+a-\rho) \\ 
					+2(s-\alpha)\\ 
					-2(s-\alpha)\\
					-(-1+s-y+a-\rho) 
  			\end{array}\right),\\  
v_4 &=& \frac{1}{4\rho}
				\left(\begin{array}{c} 
  				+(1-s+y-\alpha-\rho)\\ 
 					+2(-s+\alpha)\\ 
 					-2(-s+\alpha)\\
 					-(1-s+y-\alpha-\rho)
 				\end{array}\right),
\eeqn
and the left eigenvectors $u_{\alpha}$ are
\beq
\begin{cases}
u_1^T=(1,1,1,1) \\
u_2^T=(1,-\frac{1+y}{s+\alpha },-\frac{1+y}{s+\alpha },1) \\
u_3^T =\left(-1,\frac{2(1-y)}{1-s+y-\alpha +\rho},-\frac{2 (1-y)}{1-s+y-\alpha +\rho},1\right) \\
u_4^T =\left(-1,\frac{2 (1-y)}{1-s+y-\alpha -\rho},-\frac{2 (1-y)}{1-s+y-\alpha -\rho},1\right)
\end{cases}.
\eeq
The stationary state $P_{\infty}$ is equal to the first right eigenvector $v_1$, related to the null eigenvalue $\lambda_1$.

\subsection{Computing \texorpdfstring{$\mu$}{mu}}
\label{appB_mu}

In order to compute the effective magnetization $\mu$ in the presence of feedback, we recall Eq.~\ref{eq:Pxtz0_mu}, which is valid in general and which relates the joint probability distribution $P(x_t,z_0)$ with $\mu$. We then express $P(x_t,z_0)$ as in Eq.\ref{eq:Pxtz0_sum} in the main text:
\beq
\label{eq:appB_Pxtz0_sum}
P(x_t,z_0) = \sum_{z_t,x_0=\pm 1} P(x_t,z_t,t|x_0,z_0,0)P_0(x_0,z_0),
\eeq
where $P_0(x_0,z_0)$ is the initial distribution of the system, corresponding to the stationary state $P_{\infty}\equiv v_1(x_0,z_0)$, while the conditional probability $P(x_t,z_t,t|x_0,z_0,0)$ can be written as
\beq
\label{eq:appB_Pxtztx0z0_sum}
P(x_t,z_t,t|x_0,z_0,0) = \sum_{i=1}^{4} e^{-\lambda_i t} u_i^T(x_0,z_0) v_i(x_t,z_t).
\eeq
 $v_i$ denotes the $i$-th right eigenvector and  $u_i^T$ -- the $i$-th left eigenvector and we make the dependence on $x$ and $z$ explicit as we are going to exploit it in the subsequent algebraic manipulations. \\
We recall the definitions of $A$ and $\rho$ (Equations \ref{eq:appB_A} and \ref{eq:appB_rho}) and introduce the additional quantities
\beqn
q &=& \frac{1+y-s-\alpha}{A}, \label{eq:appB_q}\\
m &=& \frac{s+\alpha}{2A}. \label{eq:appB_m}
\eeqn
We rewrite the right eigenvectors of Appendix \ref{appB_diagL} as
\beqn
v_1(x,z) &=& \frac{1 + qxz}{4} \\
v_2(x,z) &=& mxz \\
v_3(x,z) &=& -\scriptstyle{\frac{1+y-3k+\alpha+\rho}{8\rho}} x - \scriptstyle{\frac{1+y+s-3\alpha+\rho}{8\rho}} z \\
v_4(x,z) &=& \scriptstyle{\frac{1+y-3k+\alpha-\rho}{8\rho}} x + \scriptstyle{\frac{1+y+s-3\alpha-\rho}{8\rho}} z
\eeqn
and define
\beqn
h &=& \frac{1+y}{s+\alpha}, \label{eq:appB_h}\\
a &=& \frac{1}{2} (1+s-3 y+\alpha -\rho ), \label{eq:appB_a}\\
b &=& \frac{1}{2} (-3+s+y+\alpha -\rho ), \label{eq:appB_b}\\
c &=& 1-s+y-\alpha +\rho, \label{eq:appB_c}\\
e &=& \frac{1}{2}(1+s-3 y+\alpha +\rho ), \label{eq:appB_e}\\
f &=& \frac{1}{2}(-3+s+y+\alpha +\rho ), \label{eq:appB_f}\\
g &=& 1-s+y-\alpha -\rho. \label{eq:appB_g}
\eeqn
Having done that, the left eigenvectors of Appendix \ref{appB_diagL} now read
\beqn
u_1^T(x,z) &=& 1, \\
u_2^T(x,z) &=& \frac{1-h}{2} + \frac{1+h}{2}xz, \\
u_3^T(x,z) &=& \frac{a x + b z}{c} \\
u_4^T(x,z) &=& \frac{e x + f z}{g}.
\eeqn
Now, by plugging Eq.~\ref{eq:appB_Pxtztx0z0_sum} into Eq.~\ref{eq:appB_Pxtz0_sum}, we are able to write $P(x_t,z_0)$ as
\beq
\label{eq:appB_Pxtz0_sum1}
P(x_t,z_0) = \sum_{i=1}^{4} e^{-\lambda_i t}A_i(z_0)B_i(x_t),
\eeq
where 
\beqn
A_i(z_0) &=& \sum_{x_0 \pm 1} u_i^T(x_0,z_0) v_1(x_0,z_0), \\
B_i(x_t) &=& \sum_{z_t \pm 1} v_i(x_t,z_t).
\eeqn
Computing the terms $A_i$ and $B_i$ (with $i=1,\dots,4$) we obtain: 
\beqn
A_1(z_0) &=& 1/2, \\
A_2(z_0) &=& \frac{1}{4} \left(1-h+(1+h) q z_0^2\right), \\
A_3(z_0) &=& \frac{(b+a q) z_0}{2 c}, \\
A_4(z_0) &=& \frac{(f+e q) z_0}{2 g},
\eeqn
and
\beqn
B_1(x_t) &=& 1/2, \\
B_2(x_t) &=& 0, \\
B_3(x_t) &=& -\frac{x_t (1-3 s+y+\alpha +\rho )}{4 \rho }, \\
B_4(x_t) &=& \frac{x_t (1-3 s+y+\alpha -\rho )}{4 \rho }.
\eeqn
Plugging in all the above expressions into Eq.~\ref{eq:appB_Pxtz0_sum1} we compute the effective magnetization $\mu$, which is
\beqn
\label{eq:appB_mu}
&&\mu =
\exp\left(-\frac{A}{2 \lambda}\tau \right) \Big\{q \cosh\left(\frac{\rho}{2 \lambda}\tau\right)-\notag\\
&&\frac{\left[s^2-(1+y)^2-4\alpha+\alpha^2+2s(2y+\alpha)\right]}{A\rho}\sinh\left(\frac{\rho}{2 \lambda}\tau\right)\Big\}\notag.\\
\eeqn

\subsection{Numerical results: optimal rates}
\label{appB_numerics}

\begin{figure}[!ht]
\hspace{- 0.1\linewidth}
\includegraphics[scale=0.7]{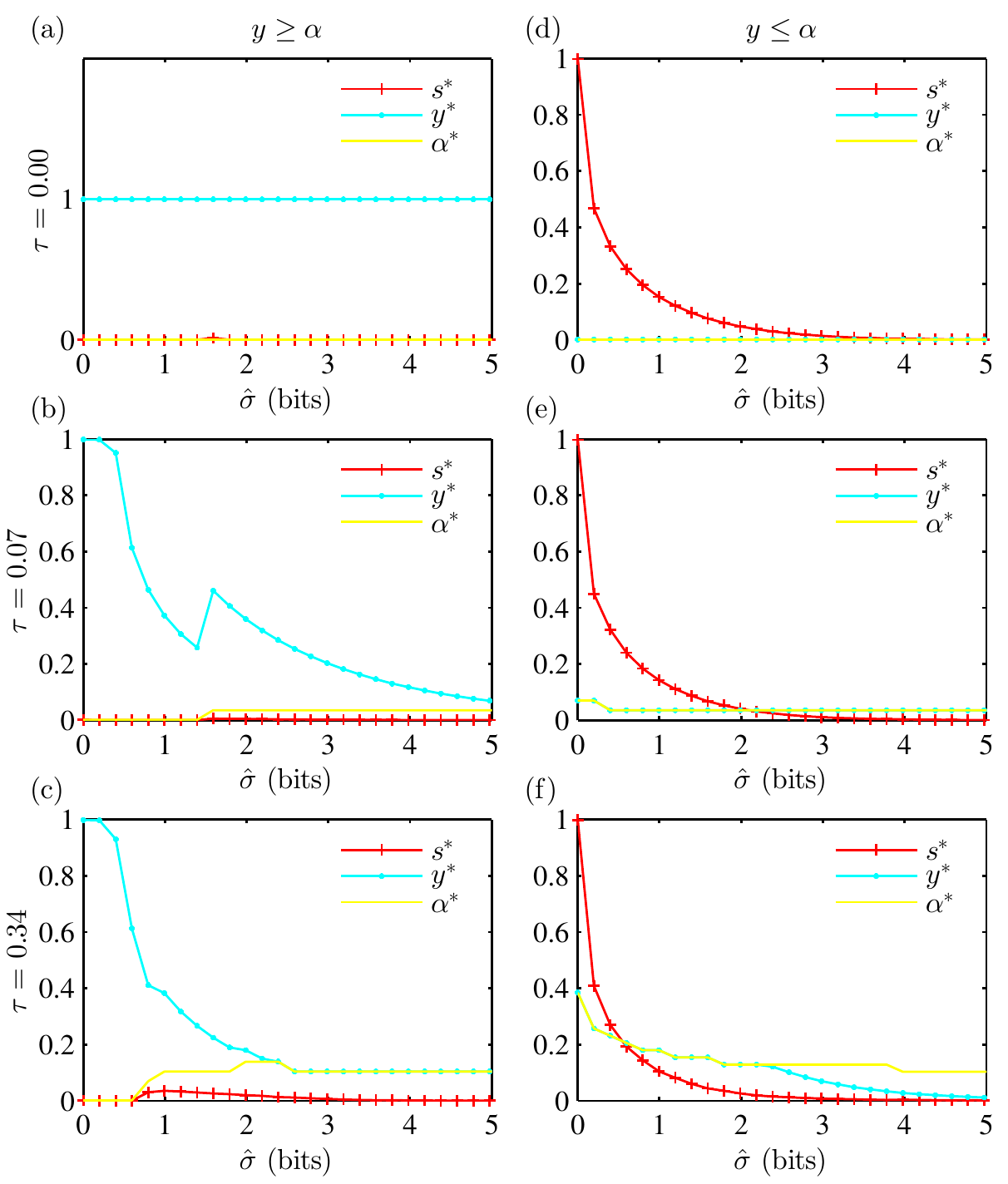}
\caption{\label{fig:panels_RatesVsDiss} Optimal rates $\{s^*,y^*,\alpha^*\}$ as functions of rescaled dissipation $\sigmahat$ for different readout delays $\tau$. Results from the simulation branch with $y\geq \alpha$ and with $y \leq \alpha$ are shown in panels (a-c) and (d-f), respectively. These rates are used to compute the optimal mutual information $\mathcal{I}^*$ of Fig.~\ref{fig:panels_MaxInfoVsDiss}a-c in the main text.}
\end{figure}
\begin{figure}[!ht]
\hspace{- 0.1\linewidth}
\includegraphics[scale=0.7]{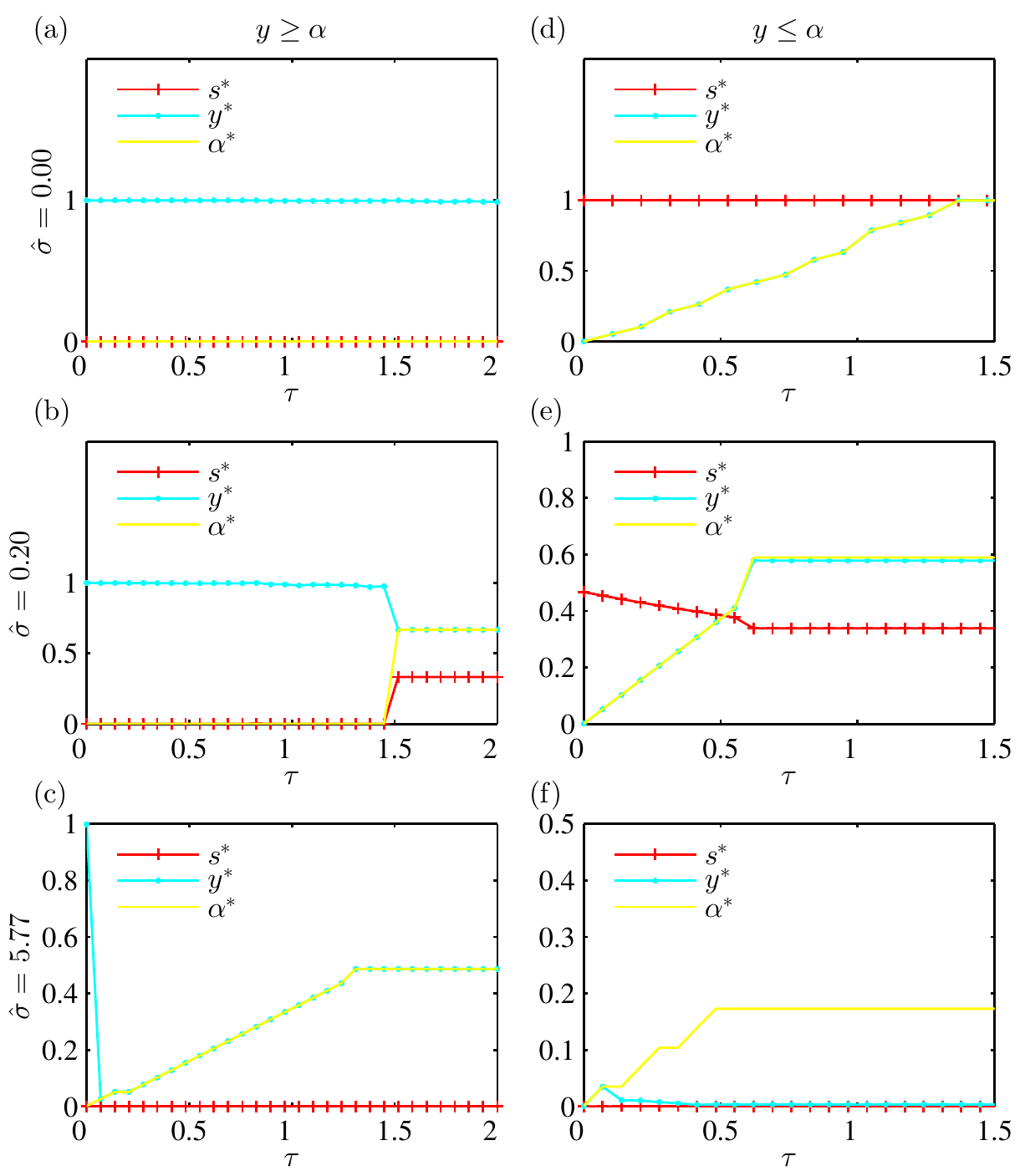}
\caption{\label{fig:panels_RatesVsTau} Optimal rates $\{s^*,y^*,\alpha^*\}$ as functions of the readout delay $\tau$ for different values of the entropy production rate $\sigmahat$ (measured in bits). Results from the simulation branch with $y\geq \alpha$ and with $y \leq \alpha$ are shown in panels (a-c) and (d-f), respectively. These rates are used to compute the optimal mutual information $\mathcal{I}^*$ of Fig.~\ref{fig:panels_MaxInfoVsTau}d-f in the main text.}
\end{figure}

In this section we show the optimal rates $\{s^*,y^*,\alpha^*\}$ resulting from numerical optimization. As discussed in the main text, optimization is performed as two separates branches: one where we fix $y\geq\alpha$, and the other where we set $y\leq \alpha$. Results from both branches are shown in Figs.~\ref{fig:panels_RatesVsDiss} and \ref{fig:panels_RatesVsTau}. \\
In the Fig.~\ref{fig:panels_RatesVsDiss}, we show the optimal rates as functions of rescaled dissipation $\sigmahat$, for different values of delay $\tau$. Such rates are used to calculate the optimal mutual information shown in Fig.~\ref{fig:panels_MaxInfoVsDiss}a-c in the main text. \\
In Fig.~\ref{fig:panels_RatesVsTau} we show the dependency of the optimal rates $\tau$, for different values of $\sigmahat$. These corresponds to $\mathcal{I}^*$ shown in Fig.~\ref{fig:panels_MaxInfoVsTau}d-f in the main text.

\section{Robust optimization}
\label{appC}

\subsection{Diagonalization of \texorpdfstring{$\mathcal{L}$}{L}}
\label{appC_diagL}
In the maximin model described in section~\ref{sec:MaxiMin}, the transition rate matrix $\mathcal{L}$ has the form
\beq
\mathcal{L} =
\left(
\begin{array}{cccc}
 u & -u & -1 & 0 \\
 -u & 1+u & 0 & 0 \\
 0 & 0 & 1+u & -u \\
 0 & -1 & -u & u
\end{array}
\right).
\eeq
Its eigenvalues $\lambda_{\alpha}$ are 
\beq
\begin{cases}
\lambda_1 = 0 \\
\lambda_2 = 1 \\
\lambda_3 = 2u \\
\lambda_4 = 1+2u
\end{cases}.
\eeq
Its right eigenvectors $v_{\alpha}$ are
\beqn
P_{\infty}=v_1 &=&\frac{1}{2(1+2u)}\left(\begin{array}{c} u+1 \\ u  \\ u  \\ u+1  \end{array}\right),\\
v_2 &=& \frac{u}{2(1-2u)}\left(\begin{array}{c} +1 \\ +1  \\ -1  \\ -1  \end{array}\right),\notag\\
v_3 &=&\frac{1}{2(1-2u)}\left(\begin{array}{c} u-1 \\ -u  \\ -u  \\ 1-u  \end{array}\right),\\
v_4 &=& \frac{u}{2(1+2u)}\left(\begin{array}{c} +1 \\ -1  \\ -1  \\ +1  \end{array}\right),
\eeqn
and its left eigenvectors $u_{\alpha}^T$ are
\beq
\begin{cases}
u_1^{T} = (1,1,1,1), \\
u_2^{T}= (-1,\frac{-u+1}{+u},\frac{+u-1}{+u},1), \\
u_3^{T} = (-1,1,-1,1),\\
u_4^{T} = (1,\frac{-u-1}{+u},\frac{-u-1}{+u},1).
\end{cases}
\eeq
The stationary state $P_{\infty}$ is equal to the first right eigenvector $v_1$, related to the null eigenvalue $\lambda_1$.

\subsection{Solution for \texorpdfstring{$\tau\ll 1$}{small tau}}
\label{appC_smalltauexp}

In this section we derive the asymptotic behavior of $u^*$ as $\tau \rightarrow 0$ given in Eq.~\ref{uasymmaximin}.
In general, Eq.~\ref{eq:condustarminimax} results in a transcendental equation for the auxiliary variable $a=\tau\frac{1-2u^*}{2u^*}$:
\beq
\label{eq:MaxiMin_eq_a}
e^{a} = 1 + a + \frac{2a^2}{\tau},
\eeq 
which needs to be solved numerically. However, in the limit $\tau\ll1$ we can analytically solve Eq.~\ref{eq:MaxiMin_eq_a}.

When $\tau\rightarrow0$, in order for the left hand side (l.h.s.) of Eq.~\ref{eq:MaxiMin_eq_a} to match the leading order term $1/\tau$ of the right hand side (r.h.s.) of Eq.~\ref{eq:MaxiMin_eq_a} , $a$ must be of the form 
\beq
a = -\log \tau + b. 
\eeq
Hence Eq.~\ref{eq:MaxiMin_eq_a} becomes
\beq
\label{eqntoplugb}
\frac{e^b}{\tau} = 1 + b -\log\tau + \frac{1}{\tau}2(b-\log\tau)^2. \notag
\eeq
Multiplying both sides by $\tau$ gives
\beq
e^b = (1+b-\log\tau)\tau + 2b^2 - 4b\log\tau +2(\log\tau)^2.\notag
\eeq
The leading order term of the r.h.s. for $\tau\rightarrow 0$ is $2(\log\tau)^2$, thus the above equation becomes
\beq
e^b \simeq 2(\log\tau)^2, \quad \tau\rightarrow 0,\notag
\eeq
which implies that $b$ has the form 
\beq
b = \log(2(\log\tau)^2) + c.
\eeq
Plugging b into Equation~\ref{eqntoplugb} we obtain 
\beqn
(2(\log\tau)^2)e^c &\simeq& 2(\log\tau)^2 -4(\log(2(\log\tau)^2)+c)\log\tau +\notag\\
&+& 2(\log(2(\log\tau)^2)+c)^2 + \dots. 
\eeqn
and by $2(\log\tau)^2$, we have
\beq
e^c = 1 - 2\frac{\log(2(\log\tau)^2)+c}{\log\tau} +\dots \simeq e^{-2\frac{\log(2(\log\tau)^2)}{\log\tau}},\notag
\eeq
which finally implies that 
\beq
c = -2\frac{\log(2(\log\tau)^2)}{\log\tau}.
\eeq
To sum up, when $\tau\ll1$ one can write $a$ as
\beqn
a&\simeq -\log\tau + \log(2(\log\tau)^2) -2\frac{\log(2(\log\tau)^2)}{\log\tau} \notag\\
& = \log\left(\frac{2(\log\tau)^2}{\tau}\right)-2\frac{\log(2(\log\tau)^2)}{\log\tau}
\eeqn
When $\tau\rightarrow 0$, $a$ diverges as
\beq
a \simeq \log\left(\frac{2(\log\tau)^2}{\tau}\right)-2\frac{\log(2(\log\tau)^2)}{\log\tau}, \quad \tau\ll1
\eeq
and $u^*$ goes to zero with $\tau$ in a strongly sublinear way:
\beq
u^* \simeq \frac{\tau}{2(-\log\tau + \log(2(\log\tau)^2) -2\frac{\log(2(\log\tau)^2)}{\log\tau})}, \tau \ll1.
\eeq

\bibliography{dissipation-information_biblio,aleks_bib}

\end{document}